\documentclass[aps,reprint,10pt,twocolumn,superscriptaddress]{revtex4-2}

\makeatletter
\def\switch@array{}
\makeatother

\usepackage{amsmath, amssymb, mathtools, bbm, braket}
\usepackage[utf8]{inputenc}
\usepackage{graphicx}
\usepackage{hyperref}
\usepackage{array}
\usepackage{booktabs}
\usepackage{multirow}
\usepackage{natbib}
\usepackage[normalem]{ulem}
\usepackage{enumitem}
\usepackage{subcaption}
\usepackage{ragged2e} 
\DeclareCaptionJustification{justified}{\justifying}

\usepackage{microtype}
\microtypecontext{spacing=nonfrench}
\microtypesetup{
protrusion={true,nocompatibility},
activate={true,nocompatibility},
tracking=true,
kerning=true,
spacing={true}
}

\newtheorem{theorem}{Theorem}
\newtheorem{definition}{Definition}

\newcommand{\eye}{\mathbbm{1}}

\newcommand{\CZ}{\mathrm{CZ}}
\newcommand{\CP}{\mathrm{C}}
\newcommand{\CCZ}[1]{\mathrm{C}^{#1}\mathrm{Z}}
\newcommand{\ccz}{\mathrm{CCZ}}
\newcommand{\CNOT}{\mathrm{CNOT}}
\newcommand{\CCNOT}[1]{\mathrm{C}^{#1}\mathrm{NOT}}

\newcommand{\LU}[1]{\mathrm{LU}_{#1}}
\newcommand{\lu}[1]{\mathfrak{lu}_{#1}}
\newcommand{\LSTAB}[1]{\mathrm{LS}_{#1}}
\newcommand{\lstab}[1]{\mathfrak{ls}_{#1}}
\newcommand{\SEP}{\mathrm{SEP}}
\newcommand{\STAB}{\mathrm{STAB}}
\newcommand{\Cliff}{\mathcal{C}}
\newcommand{\Pauli}{\mathcal{P}}
\newcommand{\States}{\mathcal{S}}
\newcommand{\Hil}{\mathcal{H}}
\newcommand{\Var}[1]{\mathrm{Var}[#1]}
\newcommand{\CC}{\mathbb{C}}
\newcommand{\Dm}{\mathcal{D}}
\newcommand{\sdiff}{\,\Delta\,}
\newcommand{\Real}{\mathbb{R}}
\newcommand{\norm}[1]{\lVert #1 \rVert}
\newcommand{\Tr}{\mathrm{Tr}}
\newcommand{\expval}[1]{\left\langle #1 \right\rangle}
\newcommand{\Comp}{K}

\renewcommand{\vec}[1]{\pmb{#1}}

\newcommand{\Avg}[1]{\langle #1 \rangle}
\newcommand{\Neg}{\mathcal{N}}
\newcommand{\floor}[1]{\lfloor #1 \rfloor}

\begin{abstract}
Quantum hypergraph states extend the well-studied class of graph states by taking into account
multi-qubit interactions through hyperedges. They provide a powerful framework to represent a family of quantum states with genuine multipartite entanglement.
In this review, we provide a compact overview of the formal structure, entanglement characteristics, and operational relevance of hypergraph states in quantum information theory. 
We begin by introducing their mathematical foundations and generalizations of the stabilizer formalism. 
A central focus is placed on their entanglement properties, including the classification under local unitary (LU) and stochastic local operations with classical communication (SLOCC), the quantification of multipartite entanglement, and detection techniques via entanglement witnesses. We also explore other nonclassical features of hypergraph states, such as contextuality and genuine multipartite nonlocality, derived from stabilizer-based Bell-type inequalities. 
Additional attention is given to the  role 
of hypergraph states in error correction, and as a computational resource in measurement-based quantum computation (MBQC), and to their non-stabilizer character - quantified via resource-theoretic measures of quantum magic.
Finally we review their generalization to higher dimensions, i.e. to qudits and continuous variables.
\end{abstract}

\begin{document}

\title{Quantum Hypergraph States: A Review}

\author{Davide Poderini}
\affiliation{Universit\`a degli Studi di Pavia, Dipartimento di Fisica, QUIT Group, via Bassi 6, 27100 Pavia, Italy}

\author{Dagmar Bruß}
\affiliation{Institut für Theoretische Physik III, Heinrich-Heine-Universität Düsseldorf, Universitätsstraße 1,
40225 Düsseldorf, Germany}

\author{Chiara Macchiavello}
\affiliation{Universit\`a degli Studi di Pavia, Dipartimento di Fisica, QUIT Group, via Bassi 6, 27100 Pavia, Italy}

\maketitle

\tableofcontents

\section{Introduction}

The study of multipartite entangled states lies at the heart of quantum information theory, with profound implications for quantum computation, quantum communication, and the understanding of nonlocal correlations. 
Among the most structurally elegant and widely applied classes of entangled states are the \emph{graph states}~\cite{raussendorf2001one, hein2004multiparty}, pure multipartite states associated with undirected graphs.

For a given mathematical graph, the corresponding graph state is defined by assigning to each vertex of the graph a qubit in a specific initial state, and applying a controlled-Z gate for each edge (i.e. a vertex subset of cardinality 
equal to $2$) of the graph. The resulting states are stabilized by commuting elements of the Pauli group~\cite{hein2004multiparty}.
Graph states are fundamental in quantum information processing, specifically in protocols such as \emph{measurement-based quantum computation} (MBQC)~\cite{raussendorf2001one, raussendorf2003measurement} and quantum error correction~\cite{schlingemann2001quantum}, but also in communication protocols, for instance for quantum secret sharing~\cite{schlingemann2001quantum}. 

However, quantum protocols and algorithms of physical and computational interest often involve interactions that go beyond pairwise couplings. 
This motivated the generalization of graph states to \emph{hypergraph states}, formally over the past decade~\cite{Quantum_hypergr_Rossi_2013, 
Encoding_en_Qu_Ri_2013, Hypergraph_stat_Rossi_2014}, which extend the graph state formalism by incorporating multi-qubit controlled-Z gates.

For a given mathematical hypergraph, the construction of
the corresponding hypergraph state resembles that of a graph state, but in this case for each hyperedge (i.e. a vertex subset of cardinality larger than or equal to $2$) a generalized controlled-Z gate is applied to all its qubits.  
The resulting quantum state encodes the combinatorial structure of the hypergraph and generalizes the stabilizer formalism associated with graph states.
However, this generalized stabilizer formalism is considerably different from the standard one based on the Pauli group.
Instead, it includes more general operators constructed from multi-qubit phase gates, and has been used as a powerful tool for the analysis of symmetries, local equivalence classes, and for entanglement detection. 
Hypergraph states are also deeply connected to other important classes of states, such as locally maximally entangleable (LME) states~\cite{kruszynska2009local}, and real equally weighted (REW) states~\cite{bruss2011multipartite}, of which in fact they are an alternative representation~\cite{qu2012quantum, Quantum_hypergr_Rossi_2013, Encoding_en_Qu_Ri_2013}.

Quantum hypergraph states exhibit a richer entanglement structure and more   intricate LU (local unitary) and  SLOCC (stochastic local operations with classical communication)  equivalence classes than graph states, and cannot in general be reduced to graph states by local Clifford operations~\cite{Entanglement_an_Guhne_2014}, showing that they represent new families of genuine multipartite entanglement.
Their classification, even for a small number of qubits, reveals a highly nontrivial structure.

Beyond structural interest, hypergraph states also offer practical advantages.
They naturally appear in certain quantum algorithms involving REW states defined by Boolean functions, such as the Deutsch–Jozsa and Grover algorithms~\cite{bruss2011multipartite, Hypergraph_stat_Rossi_2014}, and serve as universal resources for MBQC even when combined with Pauli-only measurements~\cite{Verification_of_Morima_2017,Quantum_computa_Takeuc_2019, Changing_the_ci_Gachec_2019}.
In the context of quantum error correction, \emph{hypergraph codes} have been proposed to reduce the number of entangling gates required compared to graph-state counterparts~\cite{Quantum_Error_C_Balaku_2017, Analysis_of_qua_Wagner_2018}, while still enabling fault-tolerant encoding and transversal operations. 
Moreover, hypergraph states show promise in blind quantum computation and quantum verification protocols~\cite{Efficient_Verif_Zhu_H_2019, li2023robust}, where restricted measurement bases (e.g., only $X$ and $Z$) are desirable. Randomized hypergraph state models have also been introduced~\cite{Randomized_hype_Salem_2025} to study the effects of probabilistic gate application and decoherence, providing insights into the robustness of entanglement in realistic quantum architectures.

Recent work has expanded the study of hypergraph states into domains such as non-stabilizerness (or \emph{magic})~\cite{Many_Body_Quant_Liu_Z_2022}, the key resource for quantum computational speedup, which can be effectively quantified using standard quantities like entropy of magic and stabilizer Rényi entropies~\cite{Magic_of_quantu_Chen_2024}. 
Furthermore, higher-dimensional generalizations in the form of \emph{qudit hypergraph states} have been proposed~\cite{Qudit_hypergrap_Xiong_2018,Qudit_hypergrap_Steinh_2017}, introducing a natural extension of the framework that could be relevant for novel applications in quantum coding and error correction.

Hypergraph states also offer potential applications in the field of neural networks, including the implementation of an artificial neuron using an algorithm that serves as a quantum analogue of a perceptron~\cite{tacchino2019artificial}.

Finally, hypergraph states have been shown to exhibit contextuality~\cite{Entanglement_an_Guhne_2014} and strong violations of Bell-type inequalities~\cite{Extreme_Violati_Gachec_2016}, positioning them as testbeds for foundational investigations of quantum nonclassicality.

The purpose of this review is to provide a compact overview of the main results on quantum hypergraph states, focusing in particular on their entanglement structure and properties. 
We begin with their formal definition (Sec.~\ref{sec:definitions}) and the generalized stabilizers, followed by a detailed account of their entanglement and equivalence structure (Sec.~\ref{sec:entanglement}) and their role as a resource in computation and error correction (Sec.~\ref{sec:computation}). 
Finally, we discuss higher-dimensional generalizations to qudit systems (Sec.~\ref{sec:qudits}) and continuous variables (Sec.~\ref{sec:cv}).

\section{General structure of Hypergraph States}
\label{sec:HG_intro}

\begin{figure}
    \centering
    \includegraphics[width=0.5\textwidth]{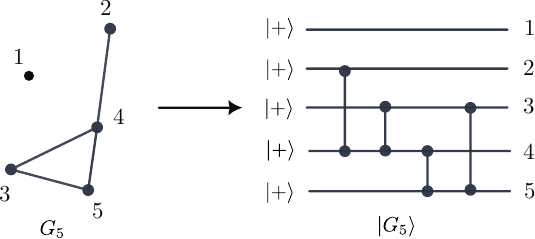}
    \caption{\textbf{Example for a graph state:} Representation of a graph state of $5$ qubits and the corresponding quantum circuit. The state is initialized as $\ket{+}^{\otimes 5}$, and a $\CZ$ gate is applied for every edge in the graph $G_5$. The $\CZ$ gates are represented by a line, with dots indicating qubits on which they are acting.}
    \label{fig:graph_state_circuit}
\end{figure}

\begin{figure}
    \centering
    \includegraphics[width=0.5\textwidth]{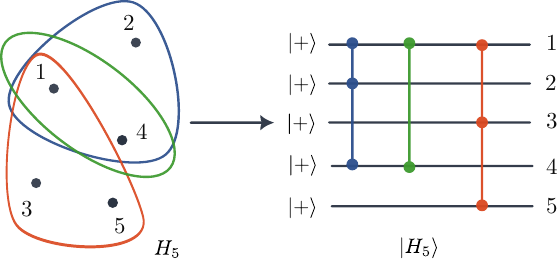}
    \caption{\textbf{Example for a hypergraph state:} Representation of a hypergraph state of $5$ qubits with edges of cardinality $2$ and $3$, and the corresponding quantum circuit. The state is initialized as $\ket{+}^{\otimes 5}$, and for every $2-$ or $3$-edge in the graph $H_5$ a $\CZ$ or $\CCZ{3}$ gate is applied, respectively. The gates are represented by a line, with dots indicating qubits on which they are acting.}
    \label{fig:hg_state_circuit}
\end{figure}

\subsection{Definitions}
\label{sec:definitions}

Hypergraph states are a direct generalization of graph states.
A graph state $\ket{G_n}$ of $n$ qubits is associated with an undirected graph $G_n = (V, E)$, where each vertex $v \in V$, with $v=1,\ldots,n$, corresponds to a qubit, initialized in the state $\ket{+}$, and each edge $\{i,j\} \in E$ indicates the application of a controlled-$Z$ ($\CZ$) gate on qubits $i$ and $j$. 
Formally, the graph state is defined as follows.
\begin{definition}[Graph state]
Given a graph $G_n = (V,E)$ on $n$ vertices the corresponding graph state $\ket{G_n}$ is given by
\begin{equation}
    \ket{G_n} = \prod_{(i,j) \in E} \CZ_{ij} \ket{+}^{\otimes n}\,,
\end{equation}
where $\ket{+} = (\ket{0} + \ket{1})/\sqrt{2}$ is the $+1$ eigenstate of the Pauli $X$ operator,
and $\CZ_{ij}= \mathrm{diag}(1,1,1,-1)_{ij}$ denotes the CZ-gate acting on qubits $i$ and $j$.
\end{definition}

Fig.~\ref{fig:graph_state_circuit} shows the quantum circuit for the generation of a graph state: first all qubits are initialized in the state $\ket{+}$, then $\CZ$ gates are applied between the qubits according to the edges in the graph.

A basic example of a graph state is the Bell-like state $\ket{\Phi} = \CZ_{12}\ket{++} = \frac{1}{\sqrt{2}}(\ket{0+} + \ket{1-})$ (albeit not written in the computational basis), which is represented as a single edge between two nodes.

\begin{figure}
    \centering    
    \begin{subfigure}{0.28\columnwidth}
        \includegraphics[width=0.7\linewidth]{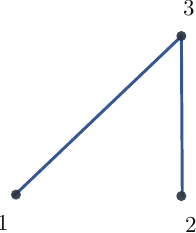}
    \caption{}
    \label{fig:hg_example_gc3} 
    \end{subfigure}
    \begin{subfigure}{0.28\columnwidth}
        \includegraphics[width=0.7\linewidth]{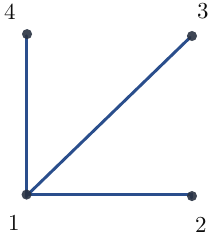}
    \caption{}
    \label{fig:hg_example_gstar} 
    \end{subfigure}
    \begin{subfigure}{0.28\columnwidth}
        \includegraphics[width=0.7\linewidth]{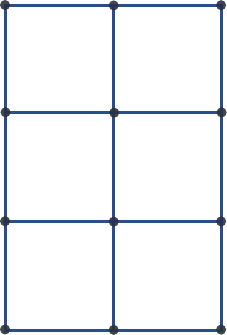}
    \caption{}
    \label{fig:hg_example_gcluster} 
    \end{subfigure}\\
    \begin{subfigure}{0.28\columnwidth}
        \includegraphics[width=0.9\linewidth]{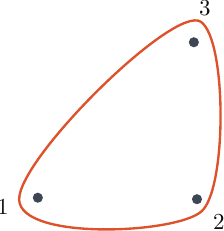}
    \caption{}
    \label{fig:hg_example_hcl3}
    \end{subfigure}    
    \begin{subfigure}{0.28\columnwidth}
        \includegraphics[width=0.9\linewidth]{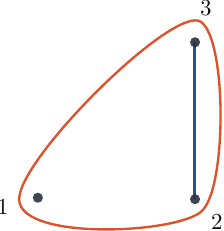}
    \caption{}
    \label{fig:hg_example_hk3_23} 
    \end{subfigure}
    \begin{subfigure}{0.28\columnwidth}
        \includegraphics[width=0.9\linewidth]{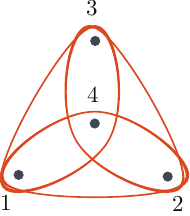}
    \caption{}
    \label{fig:hg_example_triskell} 
    \end{subfigure}\\
    \begin{subfigure}{0.28\columnwidth}
        \includegraphics[width=0.8\linewidth]{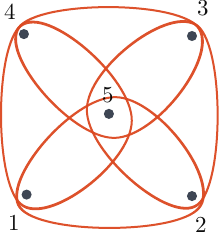}
    \caption{}
    \label{fig:hg_example_hcl5}
    \end{subfigure}    
    \begin{subfigure}{0.28\columnwidth}
        \includegraphics[width=0.8\linewidth]{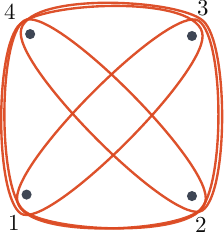}
    \caption{}
    \label{fig:hg_example_hcl4}
    \end{subfigure}
    \begin{subfigure}{0.28\columnwidth}
        \includegraphics[width=0.9\linewidth]{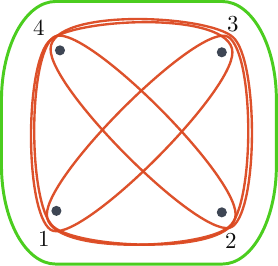}
    \caption{}
    \label{fig:hg_example_hcl34}    
    \end{subfigure}
    
    \caption{\textbf{Various examples of (hyper)graph states:}
    \subref{fig:hg_example_gc3} and \subref{fig:hg_example_gstar} examples of standard graph states on $3$ vertices, while \subref{fig:hg_example_gcluster} is a cluster state on $8$ vertices.
    \subref{fig:hg_example_hcl3} is the simplest hypergraph state, having only one global $3$-edge containing all the nodes. 
    Together with \subref{fig:hg_example_triskell} and \subref{fig:hg_example_hcl5} it is part of the family of Clover hypergraphs with $3,4$ and $5$ vertices respectively, which are $k$-uniform subgraphs sharing a vertex in all the edges.
    \subref{fig:hg_example_hcl4} is an example of a complete (or symmetric) $3$-uniform hypergraph.
    \subref{fig:hg_example_hcl34} instead is also symmetric but not $3$-uniform, since it contains also a global $4$-edge.
    }
    \label{fig:hg_examples}
\end{figure}

By adding additional nodes, we can obtain a GHZ-like state (again not in the computational basis), like the graph represented in Fig.~\ref{fig:hg_example_gc3}
\begin{multline}
    \ket{S_3} = \CZ_{13}\CZ_{23}\ket{+++} =\\
    =\frac{1}{\sqrt{2}}(\ket{++0} + \ket{--1})\,, 
\end{multline}
or the $4$-node ``star graph'' in Fig.~\ref{fig:hg_example_gstar} which corresponds to:
\begin{multline}
    \ket{S_4} = \CZ_{13}\CZ_{12}\CZ_{13}\CZ_{14}\ket{++++} =\\
    = \frac{1}{\sqrt{2}}(\ket{0+++} + \ket{1---})\,.
\end{multline}

\emph{Cluster states}, like the one represented in Fig.~\ref{fig:hg_example_gcluster} are also an important example of graph states, and represent a fundamental resource for MBQC~\cite{raussendorf2003measurement}.

A hypergraph is a natural generalization of the concept of a graph.
Similarly to the latter, we can define it as a pair $H_n=(V, E)$ of a set of $n$ vertices $V$, and a set of hyperedges $E$, where now, differently from a graph, each hyperedge $e \in E$ can be an arbitrary subset of $V$.
A hyperedge containing $k$ vertices is called $k$-edge.

Considering a hypergraph $H_n = (V, E)$ with a set of $n$ vertices $V$ and an edge set $E$, we can generalize the previous construction to a hypergraph state.

\begin{definition}[Hypergraph state]
Given a hypergraph $H_n = (V,E)$ on $n$ vertices we call hypergraph state $\ket{H_n}$ the quantum state given by
\begin{equation}
    \ket{H_n} = \prod_{e \in E} \CZ_e \ket{+}^{\otimes n},
\end{equation}
where each $k$-edge $e \subseteq V$ of cardinality $|e| = k$ corresponds to a controlled-Z operation $\CZ_e \equiv \CCZ{k} = \eye - 2\ket{1 \cdots 1}\bra{1 \cdots 1}$ that acts on the $k$ qubits which are the elements of $e$.
\end{definition}

Like for graph states, the circuit realization of hypergraph states starts by initializing all qubits in the state $\ket{+}$, but in this case we act with multicontrolled-Z gates as shown in Fig.~\ref{fig:hg_state_circuit}.

Clearly hypergraph states also include standard graph states, which are nothing but hypergraphs with only $2$-edges.
Fig.~\ref{fig:hg_example_hcl3} shows the simplest possible example of a hypergraph state (which is not also a graph state), with a single global $3$-edge on $3$ nodes, which corresponds to
\begin{equation}
    \ket{\Comp_3^3} = \CZ_{123} \ket{+++} = \frac{1}{\sqrt{8}}\sum_{i,j,k=0}^1(-1)^{ijk}\ket{i,j,k}\,.
    \label{eq:K33_compact}
\end{equation}
The effect of the gate $\CZ_{123}$ is simply to put a $-1$ phase only on $\ket{111}$ in the computational basis:
\begin{multline}
    \ket{\Comp_3^3} = \frac{1}{\sqrt{8}}
    \left(
     \ket{000}
    +\ket{001}
    +\ket{010}
    +\ket{011} +\right.\\ \left.
    +\ket{100}
    +\ket{101}
    +\ket{110}
    -\ket{111}
    \right)
    \label{eq:K33_long}
\end{multline}
Adding a $2$-edge like in Fig.~\ref{fig:hg_example_hk3_23} means applying an additional gate $\CZ_{23}$, which has the effect of moving the $-1$ to the term $\ket{011}$ instead.

A less trivial example is the one in Fig.~\ref{fig:hg_example_triskell}, which includes all the $3$-edges containing the node $4$, and can be written as:
\begin{multline}
    \ket{H_4} = \CZ_{124}\CZ_{134}\CZ_{234}\ket{++++} =\\
    = \frac{1}{\sqrt{2}}\left( \ket{+++0} + \CZ_{12}\CZ_{23}\CZ_{13}\ket{+++1}\right)\,,
\end{multline}
where we used the fact that the node $4$ acts as a ``global control''.
Noticing that the second term corresponds exactly to the complete graph state on $3$ nodes:
\begin{multline}
    \ket{\Comp_3^3} = \frac{1}{\sqrt{8}}
    \left(
     \ket{000}
    +\ket{001}
    +\ket{010}
    -\ket{011} +\right.\\ \left.
    +\ket{100}
    -\ket{101}
    -\ket{110}
    -\ket{111}
    \right)\,,
\end{multline}
we can rewrite it compactly as:
\begin{equation}
    \ket{H_4} = \frac{1}{\sqrt{2}}\left( \ket{+++0}_{1234} +\ket{\Comp_3^3}_{123}\ket{1}_4\right)\,.
\end{equation}

\paragraph*{Representation in terms of Boolean functions.}
As we will discuss also in section~\ref{sec:rew}, when expressed in the computational basis, hypergraph states have only $\pm1$ coefficients.
This can be concluded immediately from the fact that $\CZ_e$ operators can only add a $-1$ phase to the initial state $\ket{+}^{\otimes n}$.
For this reason they can also be completely characterized by the phase assignments $\pm1$ for each computational basis state $\ket{x_1,\ldots,x_n}$.
This can be done by defining a Boolean function $f(x_1,\ldots,x_n):\{0,1\}^n \to \{0,1\}$ so that $(-1)^{f(x_1,\ldots,x_n)}$ represents the corresponding phase.

This function $f$ can be explicitly constructed from the hypergraph.
For example we know that the hypergraph $\Comp_3^3$ has a single $-1$ sign on the term $\ket{111}$.
Therefore we can write $f(x_1,x_2,x_3) = x_1x_2x_3$ so that $f(111) = 1$ while $f(x_1,x_2,x_3) = 0$ for any other input.
If we now apply a $\CZ_{23}$ gate to this state we obtain the state represented in Fig.~\ref{fig:hg_example_hk3_23}.
The effect of this gate is to perform an additional flip to the phase whenever $x_2x_3 = 1$, so we can directly add this term to $f$, obtaining $f'(x_1, x_2, x_3) =  x_1x_2x_3 + x_2x_3 = \bar x_1 x_2 x_3$, where $\bar x$ denotes the Boolean NOT, and the addition is intended to be modulo $2$.
The new $f'$ now describes a state with a single $-1$ sign on the term $\ket{011}$, which is exactly what we wanted.

We can generalize this construction and define, for any hypergraph $H=(V,E)$, its corresponding Boolean function $f_H$ as:
\begin{equation}
    f_H(x_1,\ldots,x_n) = \sum_{e \in E} \prod_{j \in e} x_j\,. 
\end{equation}
We will see in section~\ref{sec:nonstab} how the properties of this function are connected with the stabilizerness of the corresponding state.

\paragraph*{Symmetric and k-uniform hypergraphs.}
A hypergraph is said to be $k$-uniform if every hyperedge contains  exactly $k$ vertices. Consequently, for the corresponding states, a $k$-uniform hypergraph state is one where all entangling operations are  $k$-qubit controlled-Z gates $\CCZ{k}$.
For $k = 2$ we recover the standard class of graph states, while for $k \ge 3$ we arrive at states where entangling gates act on three or more nodes.

A $k$-uniform hypergraph is called \emph{complete} (or alternatively \emph{fully connected}) if every possible subset of $k$ vertices forms a hyperedge.
These kinds of hypergraphs are completely defined by their number of nodes $n$ and $k$, and we will denote them by $\Comp_n^k$.
Being invariant under permutations, these hypergraphs are also called $k$-uniform \emph{symmetric} hypergraphs, and the corresponding states usually present many regularities and well-defined entanglement properties (see section~\ref{sec:entanglement}).

As an example, Fig.~\ref{fig:hg_example_hcl3}, \ref{fig:hg_example_hcl4}, and \ref{fig:hg_example_hcl5}, depict three $3$-uniform hypergraphs , with $3,4$ and $5$ vertices respectively.
Of these, only the first two are also symmetric.
The one in Fig.~\ref{fig:hg_example_hcl5} instead has the property that one vertex, vertex $5$, is shared with all the edges, a property also shared with the ones in Fig.~\ref{fig:hg_example_hcl3} and \ref{fig:hg_example_triskell}.
This makes them representatives of a family of hypergraphs sometimes called ``Clover'' hypergraphs, which are a class of $k$-uniform graphs relevant for MBQC~\cite{vigliar2021error}.

The notion of a permutation symmetric hypergraph can also be generalized to the case of non-uniformity.
In this case we denote as $\Comp^{\vec k}_n$, where the vector $\vec k = (k_1, \ldots, k_m)$ with $k_1 < k_2, \ldots < k_m$, the hypergraph which is complete for each edge cardinality $k_i$. 
For instance $\Comp^{2,3}_n$ denotes be the hypergraph on $n$ nodes that contains all possible $2$-edges and all $3$-edges on these nodes, while $\Comp^{2,3,5}_n$ would also contain all possible $5$-edges.
Another example is the graph $\Comp^{3,4}_4$, represented in Fig.~\ref{fig:hg_example_hcl34}, which is a symmetric (but non uniform) graph on $4$ vertices, and has all possible $3$-edges and the only possible $4$-edge which includes all the vertices.

An important property of states associated with symmetric hypergraphs is that their permutation symmetry enforces that the sign of the amplitude $(-1)^{f(x_1,\ldots,x_n)}$ on each computational basis state $\ket{x_1,\ldots,x_n}$ depends only on its Hamming weight $w(x_1,\ldots,x_n) = \sum_i x_i$, i.e. the number of $1$s in the bitstring $x_1,\ldots,x_n$, so that $f(x_1,\ldots,x_n) = g(w(x_1,\ldots,x_n))$.
This is because if our state is described by a symmetric hypergraph we are always free to permute the labels of the vertices without changing the resulting state.
Therefore, we must have that $f(x_1,\ldots,x_n)=f(x_{\pi(1)},\ldots,x_{\pi(n)})$ for any permutation $\pi$ of the $n$ vertices: the state must be described by a permutationally invariant function $f$.
This, in turn, means that $f$ can only depend on the number of $0$s and $1$s in the bit string, not their position, which is exactly the Hamming weight.

\subsection{Generalized Stabilizer Formalism for Hypergraph States}
\label{sec:hg_stabilizers}

The structure of graph states is closely related to the theory of stabilizer states~\cite{schlingemann2001stabilizer, gottesman1997stabilizer}, which are defined as common eigenstates with eigenvalue $+1$ of an Abelian subgroup of the Pauli group $\Pauli_n$ on $n$ qubits.
Let us call $X_i$ and $Z_j$ the corresponding Pauli operators acting on qubits $i$ and $j$ respectively.

For a graph state, we associate a stabilizer generator $K_i$ to  each qubit $i \in V$, defined as:
\begin{equation}
    K_i = X_i \prod_{j \in N(i)} Z_j,
    \label{eq:g_stab}
\end{equation}
where $N(i)$ denotes the neighborhood of vertex $i$ (i.e., the set of vertices that share an edge with $i$ in $G$).

As an example, consider the graph state represented in Fig.~\ref{fig:graph_state_circuit}, which can be written explicitly as:
\begin{equation}
    \ket{G_5} = \ket{+}_1\CZ_{24}\CZ_{34}\CZ_{35}\CZ_{45}\ket{++++}_{2345}
\end{equation}
In this case we would have $5$ stabilizers, one for each node defined by:
\begin{align*}
    &K_1 = X_1
    &K_3 = X_3 Z_4 Z_5 
    & &K_5 = X_5 Z_3 Z_4\\
    &K_2 = X_2 Z_4
    &K_4 = X_4 Z_2 Z_3 Z_5
\end{align*}
We can immediately see that $K_1\ket{G_5} = X_1\ket{G_5} = \ket{G_5}$, so $\ket{G_5}$ is indeed a $+1$ eigenstate of $K_1$.
Moreover remembering that $Z_i, Z_j, \CZ_{ij}$ commute, while $X_i\CZ_{ij} = \CZ_{ij}X_i Z_j$, one can quickly check that this is true for all $K_i$, since the various $Z_j$s operators that appear when we push $X_i$ through the $\CZ_{ij}$s are always exactly compensated in the stabilizers $K_i$.

In general, the set of all operators $\{K_i\}_{i=1}^n$ generates an Abelian subgroup $\mathcal{S}_G \subset \mathcal{P}_n$, called the \emph{stabilizer group} of the graph state. 
This group $\mathcal{S}_G$ has $2^n$ elements and uniquely stabilizes the state $\ket{G}$, i.e., $\ket{G}$ is the only state such that $K_i \ket{G} = \ket{G}$ for all $i$.

A natural question that one can ask is to what extent do hypergraph states retain the stabilizer structure of graph states, and how can it be extended.

For general hypergraph states, the stabilizer generators cannot be defined as tensor products of Pauli operators. Instead, they belong to the group of unitary operators generated by generalized controlled-Z gates~\cite{Quantum_hypergr_Rossi_2013}. 
A generalized stabilizer generator for vertex $i$ is given by:
\begin{equation}
    K_i = X_i \prod_{e \in E(i)} \CZ_{e \setminus \{i\}},
    \label{eq:hg_stab}
\end{equation}
where $E(i) = \{ e \in E \, | \, i \in e \}$ is the set of edges containing $i$, and $\CZ_{e \setminus \{i\}}$ acts on all qubits in the edge $e$ except for $i$.
As required, these operators satisfy $K_i \ket{H} = \ket{H}$ for all $i$, and commute among themselves.
Note that for a graph state Eq. (\ref{eq:hg_stab}) reduces to Eq. (\ref{eq:g_stab}).

Consider, as an example, the simple hypergraph $\ket{\Comp_3^3}$, depicted in Fig.\ref{fig:hg_example_hcl3}, with a single global $3$-edge.
The state is given by Eq.~\eqref{eq:K33_long}.
The corresponding stabilizers are:
\begin{align}
    K_1 = X_1 \CZ_{23} && K_1 = X_2 \CZ_{13} && K_3 = X_3 \CZ_{12}
\end{align}
In this case the commutation rules between $X_i$ and $\CZ_e$, when $i \in e$, generalize as:
\begin{equation}
    X_i \CZ_e = \CZ_e \CZ_{e\setminus \{i\}} X_i
    \label{eq:CCZ_X_comm}
\end{equation}
Like before, it can be easily checked that $K_i\ket{\Comp^3_3} = \ket{\Comp^3_3}$ for all $i=1,2,3$, since the additional $\CZ_{ij}$ appearing from the commutation rule~\eqref{eq:CCZ_X_comm} is compensated in $K_i$. 

For any state, the resulting Abelian group still defines a valid (in general non-Pauli) stabilizer group and uniquely identifies the state via:
\begin{equation}
    \{\ket{H}\} = \{ \ket{\psi} \in \Hil_2^{\otimes n} \, : \, K_i \ket{\psi} = \ket{\psi}, \forall i \}.
\end{equation}
where  $\Hil_2^{\otimes n}$ is the Hilbert space of $n$ qubits.
The common eigenvectors of the stabilizer group are usually called the \emph{hypergraph basis} associated to $\ket{H}$, and they can be explicitly written as
\begin{equation}
    \ket{H_I} =  \prod_{i \in I} Z_i \ket{H}.
\end{equation}
for any possible choice of $I \subseteq V$.
Moreover, analogously to a graph state, we can write a hypergraph state in terms of its stabilizers as:
\begin{equation}
    \ket{H}\bra{H} = \frac{1}{2^n} \prod_i^n \frac{K_i + \eye}{2}. 
\end{equation}

As we will see in the next section, the generalized stabilizer structure provides a powerful tool to analyze local equivalence classes, symmetry properties, and the behavior of hypergraph states under local unitary (LU) and local Clifford (LC) transformations (see section~\ref{sec:entanglement}). 
Moreover, recent studies~\cite{Symmetric_hyper_Noller_2023} have explored how symmetric hypergraph states inherit stabilizer-like features that can be exploited to analyze their entanglement properties and derive bounds on the violation of Bell inequalities.
Finally, hypergraph-state stabilizers are also making them attractive candidates for the construction of quantum error-correcting codes that can be implemented with fewer gates and have highly nontrivial structures~\cite{Quantum_Error_C_Balaku_2017, Analysis_of_qua_Wagner_2018}.

\subsection{Hypergraph States, REW and LME States}
\label{sec:rew}

An important class of quantum states, which is relevant for many quantum algorithms is the set of \emph{real equally weighted (REW)} states, defined as
\begin{equation}
    \ket{\psi_f} = \frac{1}{\sqrt{2^n}} \sum_{x=0}^{2^n - 1} (-1)^{f(x)} \ket{x}\,,
    \label{eq:REW_definition}
\end{equation}
where we used $x$ to indicate both the integer and its binary representation, running over all the $2^n$ possible bit strings $(x_1,\ldots,x_n)$, with $x_i\in \{0,1\}$, and $\ket{x} = \ket{x_1,\ldots,x_n}$ represent the corresponding state on the computational basis.
Similarly, $f: \{0,1\}^n \rightarrow \{0,1\}$ is an arbitrary Boolean function acting on the bit string represented by $x$. 
These states represent a superposition of all computational basis states with phases $\pm 1$, and naturally arise in quantum algorithms such as the Deutsch–Jozsa algorithm~\cite{deutsch1992rapid} and the Grover algorithm~\cite{grover1996fast}, after oracle calls~\cite{Hypergraph_stat_Rossi_2014}.

It can be shown that every REW state can always be written as a hypergraph state~\cite{Quantum_hypergr_Rossi_2013}, 
and vice versa. Thus, the two classes of states coincide.
Their equivalence allows a useful graphical representation of REW states, and provides a constructive framework for implementing them, using gate sequences derived from hypergraph topology.

A direct generalization of hypergraph states are the locally maximally entangleable (LME) states, introduced by Kruszynska and Kraus~\cite{kruszynska2009local}.
A pure quantum state $\ket{\psi} \in \Hil$ is said to be  LME if it can be generated from a fully separable state via the application of generalized multi-party controlled-phase operations followed by a  local unitary operator.

Formally, an $n$-qubit state $\ket{\psi}$ is LME if it admits the following representation:
\begin{equation}
\ket{\psi} = U_L \prod_{e \in E} \CP_e(\phi_e)\ket{+}^{\otimes n},
\end{equation}
where $U_L = U_1 \otimes U_2 \otimes \dots \otimes U_n$ is a tensor product of local unitaries on each qubit.
Like in the definition of hypergraph states, each $e \in E$ is a subset of the index set $\{1,2,\dots,n\}$, while the generalized controlled-phase gate
$\CP_e(\phi_e)$ is defined by:
\begin{equation}
\CP_e(\phi_e) = \eye - (1 - e^{i\phi_e}) \ket{11\dots1}_e\bra{11\dots1},
\end{equation}
with $\phi_e$ being a real phase parameter.
Note that the case $\phi_e=\pi$ corresponds to the generalized 
controlled-Z gate $\CP_e(\pi) = \CZ_e$.

\begin{figure}
    \centering
    \includegraphics[width=0.7\columnwidth]{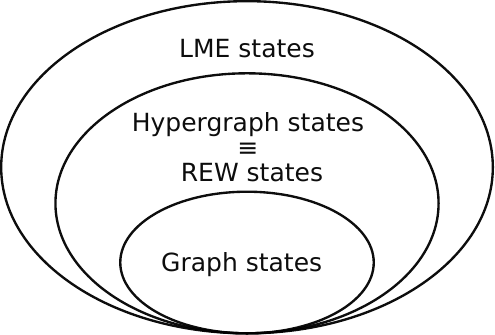}
    \caption{\textbf{Hypergraph, REW, and LME States:}}Relationship between Locally Maximally Entanglable (LME) states, hypergraph states, Real Equally Weighted (REW) states and graph states.
    \label{fig:hg_lme_rew}
\end{figure}

Hypergraph states can thus be considered as a special subclass of LME states (see Fig.~\ref{fig:hg_lme_rew}), also called $\pi$-LME states, obtained by setting $\phi_e = \pi$ for all hyperedges $e \in E$ and choosing $U_L = \eye$.
Nonetheless, LME states define a much larger class of states, presenting in general a very different entanglement structure, explicitly demonstrated by the fact that not every LME state can be transformed into a hypergraph state via local unitary operations~\cite{Relationship_am_Qu_Ri_2013}. 

\section{Entanglement properties of hypergraph states}
\label{sec:entanglement}

In this section, we present the main results concerning the entanglement properties of hypergraph states.

We have already seen that, unlike graph states, hypergraph states incorporate higher-order multipartite interactions. 
As a result, they possess a richer and more intricate entanglement structure, making them generally inequivalent to graph states under local operations, and even under the broader class of Stochastic Local Operations with Classical Communication (SLOCC)~\cite{bennett2000exact}. 

In what follows, we examine in greater detail the classification of hypergraph states under local operations, focusing on local unitary (LU) and SLOCC equivalence. 
We then turn to their genuinely multipartite entanglement features, presenting both qualitative and quantitative results. 

Different analytical and numerical tools, such as the geometric measure of entanglement~\cite{wei2003geometric} and entanglement witnesses~\cite{horodecki2001separability, terhal2000bell}, have been used to characterize and detect the entanglement properties for these states.
In particular, projector-based and stabilizer-based witnesses~\cite{Entanglement_an_Guhne_2014}, tailored to hypergraph states, have been developed as a way to study verification of genuine multipartite entanglement (GME)~\cite{horodecki2009quantum}.

Recent developments have also addressed the behavior of hypergraph states under noise~\cite{Entanglement_of_Zhou_2022, Multipartite_en_Salem_2024, Randomized_hype_Salem_2025}. Specialized purification protocols adapted to their structure demonstrate that hypergraph states can be distilled with high fidelity using only LOCC, confirming their viability as resources in realistic quantum networks and computational architectures~\cite{dur2007entanglement}.

Finally, we focus on their nonclassical behavior, with particular emphasis on contextuality and nonlocality as revealed through stabilizer-based approaches.

\subsection{LU and SLOCC Classification}

Two $n$-partite quantum states are considered to be \emph{locally unitarily} (LU) equivalent if there exists a local unitary $U_L = U_1 \otimes U_2 \otimes \dots \otimes U_n$ transforming one into the other. 

LU transformations do not affect the entanglement properties of a state. Therefore, LU-equivalence identifies a specific entanglement class. 
Each class contains states with identical entanglement characteristics, hence identifying these classes is essential for understanding the entanglement structure.

Similarly, two quantum states are said to be \emph{SLOCC equivalent}~\cite{guhne2009entanglement} if they can be transformed into one another via stochastic local operations assisted by classical communication, with nonzero probability of success.
Unlike LU equivalence, SLOCC equivalence admits non-unitary local transformations and captures a coarser classification of entanglement. 
LU-equivalent states are always SLOCC-equivalent, but the converse does not hold in general.
For instance, in the case of three qubits, it is well known that there exist exactly two inequivalent genuine tripartite entanglement classes under SLOCC: the GHZ class and the W class~\cite{dur2000three}.

A first study of LU classification for hypergraph states was  performed by Qu et al.~\cite{Multipartite_en_Qu_Ri_2013}, by analyzing the LU and SLOCC equivalence classes of three-qubit hypergraph states.
They showed that, while all such states fall into the GHZ class under SLOCC, they represent multiple distinct LU classes.
Specifically they studied $6$ LU classes, and $5$ SLOCC classes.
Notably, only one class of three-qubit hypergraph states is not LU-equivalent to any graph state.
A subsequent work~\cite{Relationship_am_Qu_Ri_2013} connected hypergraph states to the broader class of LME states. 
The authors showed that, while every hypergraph state is LME, not every LME state is equivalent to a hypergraph state, and crucially, that no $n$-qubit $W$ state is LU-equivalent to any hypergraph state. 
LU-classification quickly becomes difficult when increasing the number of qubits.
For instance, in~\cite{Locally_inequiv_Chen_2014} the authors studied the problem for $4$-qubit hypergraph states with only $2$-edges and a single global $4$-edge, already finding $11$ different LU classes.

\subsubsection{Local Pauli Operations and Their Graphical Representation}
\label{sec:graphical_ops_lu}

Among the possible local unitaries, local Pauli operations are particularly useful and manageable.
This is because they act on hypergraph states in ways that can be represented directly by a graphical transformation on the underlying hypergraph~\cite{Entanglement_an_Guhne_2014} (see table~\ref{tab:qb_ops}).

\begin{figure}
    \centering
    \includegraphics[width=0.9\linewidth]{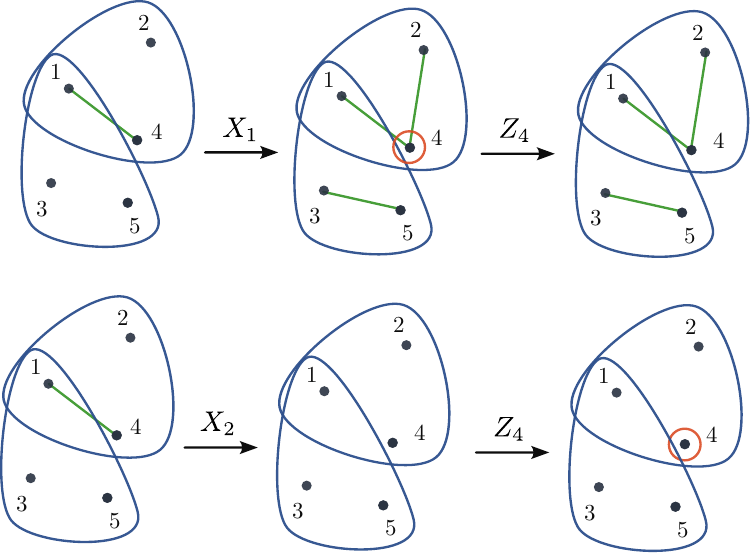}
    \caption{\textbf{Graphical Pauli rules:} Examples for the graphical rules for the action of local Pauli operators on a hypergraph state. In the top row, the action of $X_1$ on the first qubit creates two new $2$-edges, and one $1$-edge on qubit 4 since these edges are not present in the original hypergraph. 
    In the bottom row, acting with $X_2$ on qubit $2$ deletes the edge $(1,4)$ instead. Conversely the action of $Z$ can only remove or add a single $1$-edge on the index where it is applied.}
    \label{fig:pauli_graphical}
\end{figure}

Specifically the application of a local $Z_i$ gate on qubit $i$ modifies the hypergraph by \emph{adding or removing} the $1$-edge $\{i\}$ (the edge containing only the node $i$), since $Z_i = \CZ_{\{i\}}$ and $(\CZ_{\{i\}})^2 = \eye$ (see Fig.~\ref{fig:pauli_graphical}).

The application of a local $X_i$ gate instead, yields a more involved transformation:
\begin{equation}
X_i \ket{H} = \prod_{e \in E(i)} \CZ_{e \setminus \{i\}} \ket{H},
\end{equation}
Graphically, this corresponds to toggling the presence of each edge $e \setminus \{i\}$ in the hypergraph: if the edge is present, it is removed; if it is absent, it is added.
Let us define the \emph{adjacency} set of a vertex $i$ as the set of edges containing $i$ without $i$ itself, that is
\begin{equation}
    A(i) = \{e \setminus \{i\} \mid e \in E(i)\}\,.
\end{equation}
Then the new set of hyperedges simply becomes:
\begin{equation}
E \mapsto E \sdiff A(i)\,,
\label{eq:X_graphical}
\end{equation}
where $A \sdiff B$ denotes the symmetric difference between the two sets $A \sdiff B = (A\cup B) \setminus (A \cap B)$.
This action is visually intuitive: performing an $X_i$ operation induces a ``flipping'' of the participation of qubit $i$ in each edge (see Fig.~\ref{fig:pauli_graphical}).

Finally, the action of $Y_i$ can be obtained by applying both $X_i$ and $Z_i$ in sequence.`

For many generic states, local unitary equivalence reduces to local Pauli equivalence. 
That is, for some classes of states, two states are LU-equivalent if and only if they are connected by a product of local Pauli matrices~\cite{Entanglement_an_Guhne_2014}.
When possible, this simplification is very powerful and enables a purely combinatorial classification of entanglement classes based on graph-theoretic operations.
Notably, from these rules, one immediately finds that the highest-cardinality edges in a hypergraph are preserved under local Pauli operations. This already provides a simple way to distinguish inequivalent states, using the maximum edge cardinality.

Gühne et al.~\cite{Entanglement_an_Guhne_2014} investigated the local Pauli equivalence for 
$4$-qubit hypergraph states with at least one $3$-edge. 
They showed that, while for $3$ qubits only a single class exists~\cite{Multipartite_en_Qu_Ri_2013}, in the case of $4$ qubits, these states fall into $27$ local Pauli inequivalent classes.
By computing entanglement monotones like geometric entanglement and negativity (see section~\ref{sec:ent_measures}), they conclude that these represent also general LU-equivalent classes.
Moreover only three of these states exhibit maximal entanglement with respect to all bipartitions, positioning them in the so-called maximally entangled set.

\subsubsection{Edge-Pair Complementation and Permutations}

In addition to local Pauli operations, more general classes of unitary transformations, both local and nonlocal, admit a well-defined and visual interpretation in the formalism of hypergraph states. 
Two such operations, introduced in~\cite{Graphical_descr_Gachec_2017, Graph_states_an_Tsimak_2017}, are \emph{edge-pair complementation} and \emph{permutation unitaries}, which, just as local Pauli operations, preserve the hypergraph description of states (see table~\ref{tab:qb_ops}).

Local edge-pair complementation generalizes the notion of local complementation from graph states to hypergraph states. 
This operation involves a specific class of unitaries defined on a chosen qubit $i$. 
Its effect is to toggle the presence of certain \emph{edge-pairs}, i.e., unions of hyperedges that share the qubit $i$ as a common vertex.
Specifically the local edge-pair complementation at vertex $i$ modifies the hypergraph by applying the symmetric difference
\begin{equation}
E \mapsto E \sdiff \{e \cup f : e,f \in A(i)\}\,.
\end{equation}

This transformation acts on all unordered pairs of distinct hyperedges in the adjacency $A(i)$ of $i$. 
Each resulting hyperedge is the union of such a pair. 
Hence, the operation toggles edges that correspond to combinations of neighbors of $i$.
\begin{figure}
    \centering
    \includegraphics[width=0.9\linewidth]{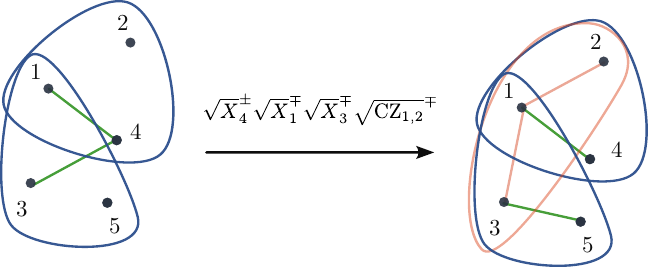}
    \caption{\textbf{Example for edge-pair local complementation}: Graphical representation of edge-pair local complementation applied on node $4$. The newly created edges represented are in red.}
    \label{fig:complementation_graphical}
\end{figure}

Physically, this transformation is implemented by a nonlocal transformation, dependent on the qubit $i$, of the form
\begin{equation}
\sqrt{X}_i^{\pm} \prod_{e \in E(i)} \sqrt{\CZ_e}^{\mp}
\label{eq:local_complementation_op}
\end{equation}
where $\sqrt{X}_i^{\pm} = \ket{+}\bra{+} \pm i\ket{-}\bra{-}$ and $\sqrt{\CZ_e}^{\pm} = \eye - (1 \mp i) \ket{1\ldots 1}\bra{1\ldots 1}$ are the square root of the Pauli $X_i$ and the control-Z operators.

While these maps can alter the entanglement properties of the state by introducing nonlocal effects, in certain hypergraph structures they can be applied to multiple vertices such that the nonlocal gates cancel out, making the overall transformation effectively local.
Notably, since the transformations that describe local complementation are not Pauli, this shows that two hypergraph states can be LU-equivalent without being equivalent for local Pauli operators.

Permutation unitaries are instead nonlocal transformations acting as permutation of the computational basis states.
On multiple qubits they are defined by an arbitrary permutation of the binary strings that label the computational basis. 
In this case, the permutation acts globally as a relabeling of the basis vectors. 
The $\CNOT$ gate and its multi-qubit controlled version $\CCNOT{k}$ are examples of this class of transformations.
On the hypergraph, the action of $\CCNOT{k}_{S,t}$, controlled by the set of qubits $S$ on the target $t$, corresponds to removing or introducing new edges defined by the union of the controls $S$ and edges containing the target.
This corresponds to taking the symmetric difference:
\begin{equation}
    E \mapsto E \sdiff \{e \cup S | e \in A(t)\}
\end{equation}
While these operations do not include all possible permutations, it is known that the set of $\CCNOT{k}_{S,t}$ operations is sufficient to construct any permutation~\cite{toffoli1980reversible}.

\begin{figure}
    \centering
    \includegraphics[width=0.9\linewidth]{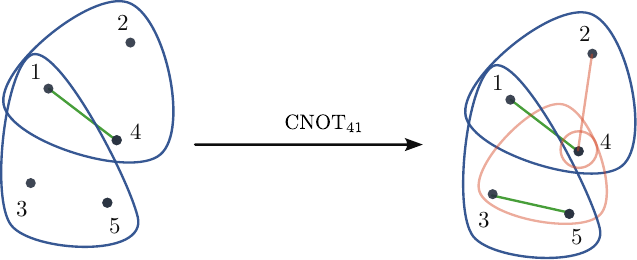}
    \caption{\textbf{Example for CNOT gate}: Graphical representation of the $\CNOT_{41}$ gate with control node $4$ and target node $1$. The newly created edges are represented in red.}
    \label{fig:CNOT_graphical}
\end{figure}

\subsubsection{Local Clifford equivalence}
\label{sec:LC_LU}

Knowing that LU-equivalence is different from general local Pauli equivalence, we could ask whether the latter can be extended to a larger  subset of unitaries which are still easier to work with than the general case.
We can introduce the notion of \emph{Local Clifford} (LC) equivalence, where the local unitaries are further constrained to be elements of the Clifford group. 
To obtain the single-qubit Clifford group we need to add to the Pauli gate set the Hadamard ($H$) and phase ($S$) gates, defined as:
\begin{equation}
    H = \frac{1}{\sqrt{2}}
    \begin{pmatrix}
        1 & 1 \\
        1 & -1
    \end{pmatrix},
    \quad
    S = 
    \begin{pmatrix}
        1 & 0 \\
        0 & i
    \end{pmatrix}.
    \label{eq:H_S}
\end{equation}
We say that two states are LC-equivalent if there exists a product of local Clifford operators $C = C_1 \otimes \cdots \otimes C_N$ that maps one into the other.

Every LC-equivalence clearly implies LU-equivalence.
It would be tempting to conjecture that LC-equivalence and LU-equivalence coincide for all cases~\cite{van2005local}.
Indeed, we know that stabilizer states have a very rigid structure that severely limits the kind of local operations that can connect them and, moreover, they are already naturally connected by Clifford operations.
This, together with the fact that the LU-LC equivalence is true for some class of graph states~\cite{van2005local, gross2007lu, zeng2007local}, led to the conjecture that this was true in general~\cite{krueger2005some}.

However, counterexamples have been found, showing that the two notions of equivalence are different in general~\cite{Graph_states_an_Tsimak_2017, Graphical_descr_Gachec_2017, ji2007lu}.
The converse, that LU-equivalence implies LC-equivalence, has nonetheless been shown to be true in some specific cases for both graph~\cite{van2005local, hein2004multiparty} and hypergraph states~\cite{Entanglement_an_Guhne_2014, Locally_inequiv_Chen_2014}, but only for small number of qubits, specifically for states with $n\le8$ and $n\le 4$ qubits, respectively.

\subsubsection{Local Stabilizers}
\label{sec:local_stab}

Another tool to investigate LU equivalence among multipartite quantum states is the analysis of their \emph{local stabilizer} subgroup~\cite{Local_unitary_s_Lyons_2015, Local_Pauli_sta_Lyons_2017}.
This method allows the study of LU equivalence of states through their infinitesimal symmetries analyzing the Lie algebraic structure of the local stabilizers subgroup.

Given a pure state $\ket{\psi}$ on $n$ qubits, one defines its \emph{local unitary stabilizer group} as
\begin{equation}
\LSTAB{\psi} = \left\{ U \in \LU{n} \, \mid \, U\ket{\psi} = \ket{\psi} \right\}.
\end{equation}
where $\LU{n}$ represents the local unitary group.
To study this group at the infinitesimal level, one considers the corresponding Lie algebra $\lstab{\psi}$.
This is the subalgebra of the local unitary group algebra $\lu(n)$, formed by zero trace skew-Hermitian operators such that every $U \in \LSTAB{\psi}$ can be expressed by exponentiation of one of its elements $U(\theta) = e^{\theta X} \quad X\in\lstab{\psi}$.
Notice that the algebra $\lstab{\psi}$ can also be written as:

\begin{equation}
\lstab{\psi} = \left\{ X \in \lu{n} \, \mid \, X\ket{\psi} = 0 \right\}\,.
\end{equation}

The stabilizer subalgebra encodes all continuous local symmetries of the state, and it is invariant under local unitaries.

Specifically if $\ket{\psi}$ and $\ket{\phi}$ are LU equivalent then there exist some $U \in \LU{n}$ such that $\ket{\psi} = U\ket{\phi}$. Hence for any $V \in \LSTAB{\phi}$ we have $UVU^\dagger\ket{\psi} = \ket{\psi}$, that is $UVU^\dagger \in \LSTAB{\psi}$, and conversely if $W \in \LSTAB{\psi}$ then $U^\dagger WU \in \LSTAB{\phi}$.

Therefore, if two states are LU-equivalent, their LU stabilizer group and subalgebras must be isomorphic:

\begin{equation}
    \LSTAB{\psi} = U\; \LSTAB{\phi} \;U^\dagger \quad
    \lstab{\psi} = U\; \lstab{\phi} \;U^\dagger\,.
\end{equation}

This necessary condition provides an efficient way to distinguish inequivalent states: if the stabilizer subalgebras differ, the states cannot be LU-equivalent. Moreover, in some structured families of states—such as graph or symmetric hypergraph states—this criterion can also be sufficient.
In practice, $\lu{\psi}$ is computed by solving the equation $X\ket{\psi} = 0$ where $X$ ranges over traceless skew-Hermitian matrices acting locally.

An important result~\cite{Local_unitary_s_Lyons_2015} is that for $k$-uniform hypergraph states the stabilizer subalgebra simplifies considerably and its elements act non-trivially only on a subset $\hat V \subseteq V$ of \emph{essential vertices}. 
This concept identifies a subset of vertices $i \in \hat V$ such that for any $e \in E(i)$, the reduced edge $e \setminus \{i\}$ is also the reduced edge of another vertex, i.e. there exist $j$ and $f \in E(j)$ for which $f \setminus \{j\} = e \setminus \{i\}$. This relationship also define \emph{essential edges} $\hat E$, which in turn can be used to construct the \emph{essential subgraph} $\hat H$.
This concept allows one to simplify the hypergraph by discarding edges that are LU-trivial and can be used as a canonical representative of the LU-equivalence class, since two hypergraph states with non-isomorphic essential graphs cannot be LU-equivalent.
This notion gives a minimal description of the entanglement structure of a hypergraph state and provides a systematic way to reduce redundancy to classifying states up to LU-equivalence.

Remarkably, for symmetric hypergraph states, one can derive necessary and sufficient conditions for the existence of local stabilizers~\cite{Local_Pauli_sta_Lyons_2017}, called \emph{palindrome conditions}.
These conditions are valid for  general non-uniform symmetric hypergraph states $\Comp^{\vec k}_n$, where $\vec k = (k_1, \ldots, k_m)$ was defined in section~\ref{sec:definitions}.
We recall that, due to their symmetry, the amplitudes in the computational basis for this class of states are determined only as a function of the Hamming weight $w(s)$ of the corresponding bitstring $s$.
If we define
\begin{equation}
    h(u) = \sum_{i=1}^m \binom{u}{k_i},
\end{equation}
it is possible to prove~\cite{Local_Pauli_sta_Lyons_2017} that a symmetric state $\ket{\Comp^{\vec k}_n}$ is: 
\begin{itemize}
    \item Stabilized by $X^{\otimes n}$ if and only if $h(w) = h(n-w) \mod 2$ for any $0\le w \le n$.
    \item Stabilized by $-X^{\otimes n}$ if and only if $h(w) = h(n-w) + 1 \mod 2$ for any $0\le w \le n$.
    \item Stabilized by $Y^{\otimes n}$ if and only if $h(w) = h(n-w) +w + n/2 \mod 2$ for any $0\le w \le n$.
\end{itemize}
The name \emph{palindrome conditions} comes from the fact that for even $k$ and $n \equiv 2 \pmod{4}$, a symmetric $k$-uniform hypergraph state is stabilized by $X^{\otimes n}$ if and only if the Boolean function $g(w)$ defining its amplitude signs is a \emph{palindrome}, i.e. $g(w) = g(N - w)$ for all  $w \in \{0,1,\dots,n\}$.
This ensures that the sign pattern of the state amplitudes in the computational basis is symmetric under bitwise inversion, which is equivalent to invariance under global $X$ operations.
One can also show~\cite{Local_Pauli_sta_Lyons_2017} that these are the only possible (nontrivial) local Pauli stabilizers for a symmetric state associated with $\Comp^{\vec k}_n$.

\subsection{Quantification and Detection of Multipartite Entanglement}
\label{sec:ent_measures}

Entanglement in multipartite systems is considerably richer and more complex than in bipartite settings~\cite{horodecki2009quantum}. 
Unlike the bipartite case, one can separate entangled states in \emph{biseparable}, and \emph{genuinely multipartite entangled} (GME) states~\cite{jungnitsch2011taming, acin2001classification}.
An $n$-partite \emph{pure} state is called \emph{biseparable} if it is separable with respect to some bipartition $\{A,B\}$, where $A = \{1, \ldots, k\}$ and $B = \{k+1, \ldots, n\}$ form an arbitrary partition of the $n$ qubits. 
If a pure state is not biseparable across any bipartition it is said to be GME, that is, its entanglement cannot be described by separability in any split of the system.
\emph{Mixed} states are classified similarly, with the specification that a biseparable mixed state is a convex combination of projectors onto pure states, each being separable across (potentially different) bipartitions.

\subsubsection{Geometric entanglement}
\label{sec:geom_ent}

An important quantity that has been introduced to characterize entanglement is the \emph{geometric entanglement}.
For a pure state $\ket{\psi}$  it is defined as:
\begin{equation}
  E_G(\ket{\psi}) = 1 - \max_{\ket{\phi} \in \SEP} |\braket{\phi | \psi}|^2,
  \label{eq:geom_ent}
\end{equation}
where the maximum is taken over all fully separable states $\ket{\phi}$.
This quantity, closely linked to the \emph{Relative Entropy of Entanglement}~\cite{vedral1997quantifying}, has been extensively studied for graph states, and analytical results are known for several classes of graph states, such as cluster, star and ring states~\cite{hajduvsek2013direct}.

For the case of GME, it can be generalized as
\begin{equation}
  E_{M}(\ket{\psi}) = \min_{\{A,B\} \in \mathcal{B}} E_G^{A|B}(\ket{\psi}),
  \label{eq:geom_GME}
\end{equation}
where the minimum is taken on the set of all bipartitions $\mathcal{B}$, and $E_G^{A|B}$ is defined as in~\eqref{eq:geom_ent}, but for a specific bipartition $\{A, B\}$:
\begin{equation}
  E_G^{A|B}(\ket{\psi}) = 1 - \max_{\ket{\phi} \in \SEP_{A|B}} |\braket{\phi|\psi}|^2.
  \label{eq:geom_ent_bip}
\end{equation}

In~\cite{Multipartite_en_Ghio_2018} the authors compute this quantity for symmetric hypergraph states with uniform edge cardinality of $n$ and $n-1$.

To illustrate the idea consider the case of $\ket{\Comp_3^3}$, i.e.
a 3-qubit hypergraph state with a single $3$-edge shown in Fig.\ref{fig:hg_example_hcl3}.
In the computational basis, this correspond to the state in Eq.~\eqref{eq:K33_long}, where all terms are positive except $\ket{111}$ which carries a negative sign.
Bipartitions in this case are equivalent due to the state's permutation symmetry. 
Choosing $A=\{1,2\}$ and $B=\{3\}$ the Schmidt decomposition yields:
\begin{equation}
\ket{\Comp^3_3} = \sqrt{\frac{3}{4}} \left( \frac{\ket{00} + \ket{01} + \ket{10}}{\sqrt{3}} \right) \ket{+} + \sqrt{\frac{1}{4}} \ket{11} \ket{-}
\end{equation}
The maximum squared Schmidt coefficient is $(s_{\max})^2 = 3/4$, which represents also the maximum overlap on separable states, therefore we have
\begin{equation}
E(\ket{\Comp_3^3} = 1 - \frac{3}{4} = \frac{1}{4}\,.
\end{equation}

\begin{table}[htpb]
    \centering
    \begin{tabular}{llc}
        State & Condition & $E_M(\ket{H})$ \\
        \toprule
        $H_n^n$ & All $n \ge 3$ & $\frac{1}{2^{n-1}}$\\
        
        \midrule
        \multirow{3}{*}{$H_n^{n-1}$} & $n = 4$ & $\frac{5-\sqrt{5}}{8}$ \\
         & $n \ge 6$, even & $\frac{n}{2^{n-1}}$ \\
        & $n \ge 3$, odd & $\frac{n-1}{2^{n-1}}$ \\
        
        \midrule
        \multirow{3}{*}{$H_n^{n-1,n}$} & $n = 3$ & $\frac{1}{4}$ \\
        & $n \ge 4$, even & $\frac{n-1}{2^{n-1}}$ \\
        & $n \ge 5$, odd  & $\frac{n}{2^{n-1}}$ \\
        \bottomrule
    \end{tabular}
    
    \caption{Summary of analytical results for multipartite entanglement ($E_M$) for specific families of $n$-qubit symmetric hypergraph states with $n-1$ and/or $n$ edge cardinalities.}
    \label{tab:hypergraph_entanglement}
\end{table}

It can be proved that this is a particular instance of a general result for hypergraph states $\ket{\Comp^n_n}$ with exactly one $n$-hyperedge, which can be further generalized for states $\ket{\Comp_n^{n-1,n}}$, additionally containing all possible hyperedges of cardinality $n-1$, and $\ket{\Comp_n^{n-1}}$ including all $n-1$ hyperedges only.
The results are summarized in Table~\ref{tab:hypergraph_entanglement}.

More generally a lower bound can be found for hypergraph states $\ket{H_{n,k}}$, where $H_{n,k}$ denotes a hypergraph on $n$ nodes with maximum hyperedge cardinality $k$, specifically:
\begin{equation}
     E_{M}(\ket{H_{n,k}}) \ge \frac{1}{2^{k-1}}\,,
\end{equation}
whose minimum value $\frac{1}{2^{k-1}}$ can be reached, as we have seen, for $k=n$.

Regarding the quantity $E_G$, as defined in Eq.~\eqref{eq:geom_ent}, a significant simplification arises for hypergraph states where one vertex is present in all hyperedges~\cite{Symmetric_hyper_Noller_2023}, like Clover graphs mentioned in section~\ref{sec:definitions}.
In this case, the state can be written in the form:
\begin{equation}
    \ket{H_n} = \frac{1}{\sqrt{2}} \left( \ket{0} \ket{+}^{\otimes (n-1)} + \ket{1} \ket{\tilde{H}_{n-1}} \right),
    \label{eq:Hn_decomposition}
\end{equation}
where $\ket{\tilde{H}_{n-1}}$ is a hypergraph state on the remaining $n-1$ qubits. 
By applying a Hadamard transformation on the distinguished qubit, one can always obtain a state with real, positive amplitudes in the computational basis, which allows $E_G$ to be computed using real product vectors only~\cite{Symmetric_hyper_Noller_2023}, which makes the analysis considerably easier.
For example, consider again the state $\ket{\Comp_n^n}$. 
To find $E_G(\Comp_n^n)$, we need to evaluate its maximum overlap with a symmetric, product state of the form $(a\ket{0}+b\ket{1})^{\otimes n}$, with $|a|^2 + |b|^2 = 1$, which is $\frac{1}{\sqrt{2}^{n}}((a+b)^{n}-2b^{n})$.
Since there is only one hyperedge, $\ket{\Comp_n^n}$ trivially satisfies the requirements for the decomposition~\eqref{eq:Hn_decomposition}:
\begin{equation}
\ket{\Comp_{n}^{n}} = \frac{1}{\sqrt{2}} \left( \ket{0}\ket{+}^{\otimes n-1} + \ket{1}\ket{\Comp_{n-1}^{n-1}} \right)
\end{equation}
Therefore we can restrict to real-valued coefficients $a$ and $b$ (with $a^2 + b^2 = 1$), obtaining:
\begin{equation}
E_{G}(\ket{H_{n}^{n}}) = 1 - \frac{1}{2^{n}} \max_{a^{2}+b^{2}=1} ((a+b)^{n} - 2b^{n})^{2}\,.
\label{eq:EG_Hnn}
\end{equation}
Evaluating this quantity as $n$ grows they obtain, as expected, that the entanglement rapidly diminishes.

Extending this idea, one can develop, for symmetric hypergraph states, a systematic method to compute $E_G$ analytically. 
If a hypergraph state is stabilized by a local Pauli operator $P^{\otimes n}$ (with $P = X$ or $Y$), then, under certain conditions on the parity function $f(w)$ (as defined in Eq.~\eqref{eq:REW_definition}), it can be mapped to a state with only real, positive coefficients.
This can be done via application of the local unitary $({\sqrt{P}^+})^{\otimes n}$, where $\sqrt{P}^+$ denotes the square root of the operator $P$ with eigenvalues ${1, i}$.
For instance, if $P=Y$, $n = 0 \mod 2^{r-1}$, and when $f$ is $2r$-periodic and $f(w) = 0$ for all even $w$, the transformed state $\ket{\tilde{\Comp}^k_n} = {(\sqrt{Y}^+)}^{\otimes n} \ket{\Comp_n^k}$ becomes a superposition of the GHZ state and a state with support only on odd-weight basis vectors:
\begin{equation}
    \ket{\tilde{\Comp}^k_n} = \frac{1}{\sqrt{2}} \left( \ket{\mathrm{GHZ}} + \ket{\psi_{\mathrm{odd}}} \right).
    \label{eq:GHZ_odd_decomposition}
\end{equation}
Similar results can be obtained for $k$-uniform hypergraph states $H_n^k$, with $k=(2^{r-1}+1)$, for a specific number of qubits, that have $X^{\otimes n}$ or $Y^{\otimes n}$ as stabilizers.
Using these results, it is possible to find an explicit form, or at least bounds, for $E_G$ in several cases.
For instance, considering uniform hypergraphs with 3-edges $H_n^3$ we have
\begin{equation}
    E_G(\ket{H^3_n}) = \frac{3}{4} - \frac{1}{2^{n/2}} - \frac{1}{2^{n}}\, ,
\end{equation}
for $n=2 \mod4$ and $\ket{H^3_n}$ is stabilized by $X^{\otimes n}$, while
\begin{equation}
    \frac{3}{4} - \frac{1}{2^{n/2}} - \frac{1}{2^{n}} \le E_G(\ket{H^3_n}) \le \frac{3}{4} - \frac{1}{2^{n}}\, ,
\end{equation}
when $n=0 \mod 4$ for which the state is stabilized by $Y^{\otimes n}$.

\subsubsection{Negativity and other entanglement measures for hypergraph states}

Other quantities like the \emph{purity} and the \emph{negativity}~\cite{plenio2005logarithmic} 
w.r.t. a bipartition, and the \emph{concurrence}~\cite{wootters1998concurrence} are sometimes appropriate to capture multipartite entanglement.

\paragraph*{Purity:}
The \emph{purity} of a quantum state $\rho$  is defined as
$P(\rho) = \mathrm{Tr}(\rho^2)$.
For a given bipartition $\{A, B\}$ of a pure global state $\ket{\psi_{AB}}$, with the reduced state  
$\rho_A = \mathrm{Tr}_{B}(\ket{\psi_{AB}}\bra{\psi_{AB}})$,  its purity $P_A = \mathrm{Tr}(\rho_A^2)$ 
quantifies the amount of entanglement between $A$ and $B$.
This quantity can also be associated with the Rényi entropy of order 2~\cite{renyi1961measures} given by
\begin{equation}
S_2(\rho_A) = -\log_2 \mathrm{Tr}(\rho_A^2) = -\log_2 P_A,
\end{equation}
which quantifies the entanglement entropy across the bipartition.

\paragraph*{Concurrence:}
For a general mixed state $\rho$ of two qubits, an alternative measure is the concurrence~\cite{wootters1998concurrence}, defined as
\begin{equation}
C(\rho) = \max\left\{0, \lambda_1 - \lambda_2 - \lambda_3 - \lambda_4 \right\},
\end{equation}
where $\{\lambda_i\}$ are the square roots of the eigenvalues, in decreasing order, of the non-Hermitian matrix
\begin{equation}
R = \rho \tilde{\rho}, \qquad \tilde{\rho} = (\sigma_y \otimes \sigma_y) \rho^* (\sigma_y \otimes \sigma_y),
\end{equation}
where $\rho^*$ is the complex conjugate of $\rho$ in the computational basis. 
A generalization of the concurrence for multi-qubit states has been suggested in~\cite{carvalho2004}.

\paragraph*{Negativity:}
A general and widely used entanglement monotone to quantify bipartite entanglement in a mixed state is the \emph{negativity}, which is based on the Peres–Horodecki criterion which relies on the partial transposition~\cite{peres1996separability,horodecki2009quantum}.
The criterion is based on the fact that a separable state $\rho_{AB}$ remains positive under partial transposition applied to one of its subsystems.
Let $\rho_{AB}^{T_A}$ denote the partial transpose of $\rho_{AB}$ with respect to subsystem $A$.
Then if $\rho_{AB}^{T_A}$ is not positive semi-definite we can conclude that $\rho_{AB}$ is entangled.
The converse however is famously not true in general, except for qubits and systems composed of a qubit and a qutrit, hence, there are states with positive partial transpose (PPT) which are nonetheless entangled.

The negativity of $\rho_{AB}$ is defined as
\begin{equation}
\Neg(\rho_{AB}) = \frac{\norm{\rho^{T_A}_{AB}}_1 - 1}{2},
\end{equation}
where $\norm{X}_1 = \Tr(\sqrt{X^\dagger X})$ denotes the trace norm.

Clearly, in general, $\Tr(X) \le \norm{X}_1$, where the equality is valid iff $X$ is positive semi-definite.

Therefore, since $\Tr(\rho_{AB}) = \Tr(\rho_{AB}^{T_A})  = 1$, the negativity vanishes for all PPT states, which includes separable states, and is strictly positive for entangled states with non-positive partial transpose (NPT).

For an $n$-partite mixed state $\rho$ this notion can be generalized to \emph{genuine multipartite negativity} (GMN)~\cite{jungnitsch2011taming}.
If $\mathcal{B}$ denotes, as before, the set of all bipartitions $\{A,B\}$ of the system, for each such bipartition, we can define the bipartite negativity $\Neg_{A|B}(\rho)$. 
The GMN is then defined as
\begin{equation}
\Neg_{\mathrm{GM}}(\rho) = \min_{\{A,B\} \in \mathcal{B}} \Neg_{A|B}(\rho).
\end{equation}
This measure is strictly positive only if the state is NPT across every bipartition and thus GME.
However, similarly to the bipartite case, this criterion cannot detect GME for PPT entangled states.

These monotones are suitable for studying the entanglement of mixed states, like those arising from randomized or noisy constructions of hypergraph states, described in section~\ref{sec:rand_hg}.

\subsubsection{Entanglement Witnesses}

An alternative tool for the characterization of the entanglement are \emph{entanglement witnesses}~\cite{horodecki2009quantum}, which have the advantage of also providing experimentally accessible tools to detect entanglement.

Formally a Hermitian operator $W$ is an entanglement (GME) witness if $\Tr(W\rho) \geq 0$ for all separable (biseparable) states $\rho$, and there exists some entangled (GME) state $\sigma$ such that $\Tr(W\sigma) < 0$.

For hypergraph states, two main classes of entanglement witnesses have been considered~\cite{Multipartite_en_Ghio_2018} which we can classify as \emph{projection-based} and \emph{stabilizer-based}.
The projection-based witness, introduced for 3-qubit states in~\cite{Entanglement_an_Guhne_2014} and generalized in~\cite{Multipartite_en_Ghio_2018}, is defined directly from the target hypergraph state $\ket{H}$ and takes the form
\begin{equation}
    W_P = \alpha \cdot \eye - \ket{H}\bra{H},
    \label{eq:proj_W}
\end{equation}
where $\alpha = 1-E_M(\ket{H})$ is the maximal squared overlap between $\ket{H}$ and any biseparable pure state, see Eq. (\ref{eq:geom_GME}).
This type of witness is conceptually simple and optimal in the sense that it detects all states with fidelity greater than $\alpha$ with respect to $\ket{H}$.
A suitable choice of the parameter $\alpha$ for a generic state of $n$ qubits, with maximum edge cardinality $k$ is $\alpha = \frac{2^{k-1} -2}{2^{k-1}}$.
However, implementing the projection $\ket{H}\bra{H}$ locally generally requires a large number of measurement settings, possibly as high as $\frac{3^n-1}{2}$, scaling exponentially with the number of qubits.

An alternative GME witness, experimentally more accessible, can be constructed using the stabilizer generators $\{K_i\}_i$ of the hypergraph state, defined in Eq. (\ref{eq:hg_stab}):
\begin{equation}
W_S = \beta \cdot \eye - \sum_{i=1}^n K_i,
\end{equation}
where $\beta$ is a constant chosen such that $W_S$ is positive on all biseparable states. This witness typically requires fewer local measurement settings, scaling linearly with the number of qubits. Moreover, $W_S$ is more accessible experimentally due to its decomposability into tensor products of Pauli and $\CCZ{k}$ gates, though it may be weaker in detecting entanglement for states not close to $\ket{H}$.

In~\cite{Multipartite_en_Ghio_2018} the authors obtained the value of $\beta$ for several families of hypergraph states, with uniform edge cardinality and highly symmetric structures.
A suitable choice for the  parameter $\beta$ for a generic hypergraph state of $n$ qubits with maximum edge cardinality $k$ is $\beta = \frac{n 2^{k-1} -2}{2^{k-1}}$.

\subsection{Entanglement Purification}
Entanglement purification~\cite{dur2007entanglement} is a protocol that aims to distill high-fidelity entangled states from multiple copies of noisy, imperfectly entangled quantum states. 
The process relies exclusively on local operations and classical communication (LOCC) between the parties.

Given several copies of an entangled state that has been degraded by interaction with a noisy environment, a purification protocol probabilistically transforms these into fewer copies of a target pure entangled state with higher fidelity.

Recently in~\cite{Entanglement_pu_Vandre_2023} the authors developed purification protocols tailored to hypergraph states. Building on earlier methods for LME states, they introduced adaptive and optimized purification procedures based on the graphical representation of the state. Their approach improves the noise threshold and fidelity of purification.

In this protocol purification success is correlated with the hyperedge structure: states with larger or more frequent hyperedges typically exhibit higher entanglement and require more sophisticated strategies. 
Nonetheless, the protocols succeed in distilling high-fidelity hypergraph states from noisy ensembles using only local operations and classical communication (LOCC).

\subsection{Entanglement properties of random hypergraph states}
\label{sec:rand_hg}

Random hypergraph states (RHS) extend standard hypergraph states to model multipartite entanglement under conditions of randomness or noise.

Under these conditions we no longer assume that the gates $\CZ_e$ are ideal, but instead only succeed with a certain probability $p_{|e|}$, which can in general depend on the edge cardinality $|e|$.
This generates a state which is a mixture of hypergraph states.

A way to model this~\cite{Multipartite_en_Salem_2024}, inspired by similar works on graph states~\cite{wu2014randomized}, is to introduce probabilistic gates $\Lambda_e(p_{|e|})$,  which apply either $\CZ_e$ or the identity with probability $p_{|e|}$, in place of the ideal control-Z operations.

The effect of these noisy gates can be defined by introducing a randomization operator $R_P$ acting on a pure hypergraph state $\ket{H}$, where $P = \{p_k\}_{k=1}^n$ is the success probability of applying a gate $\CZ_e$ for a hyperedge of order $k=|e|$.
The resulting state can be written as:
\begin{equation}
R_P(\ket{H}) = \sum_{F \preceq H} \left( \prod_{e \in E_F} p_{|e|} \prod_{e \in E_H \setminus E_F} \left(1 - p_{|e|}\right) \right) \ket{F}\bra{F}.
\end{equation}
where $F$ ranges over the spanning sub-hypergraphs $F \preceq H$.
Recall that for a given $H=(V,E)$ we call a spanning sub-hypergraph a hypergraph $F = (V',E')$ such that $V=V'$ and $E' \subseteq E$, i.e., $F$ includes all vertices of $H$ but possibly fewer edges.
This yields a mixed state whose entanglement depends on the full spectrum of edge cardinalities and their respective noise levels.

\begin{figure}
    \centering
    \includegraphics[width=0.7\linewidth]{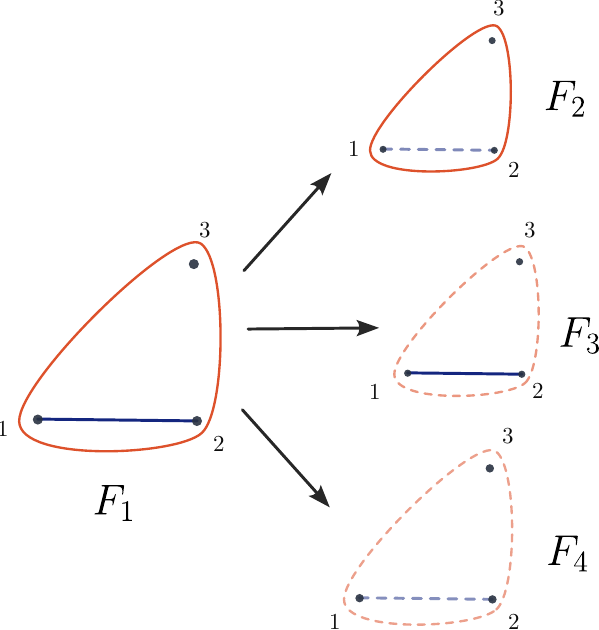}
    \caption{\textbf{Random hypergraph:} Visualization of the effect of the randomization operator on a simple hypergraph.
    The resulting state is a mixture of all the hypergaphs $F_1,\ldots,F_4$, obtained by removing one by one the edges from the original hypergraph.  
    }
    \label{fig:random_hg}
\end{figure}

As an example consider the graph in Fig.~\ref{fig:random_hg}.
In the picture the original graph $F_1$ is represented with all its spanning subgraphs $F_2,F_3,F_4$.
The action of the operator $R_P$, with $P = \{p_2, p_3\}$ results in a mixture of all the hypergaphs $F_1,\ldots,F_4$:
\begin{multline}
    R_P(\ket{H}) = \Lambda_{12}(p_2)\circ\Lambda_{123}(p_3) \ket{+}\bra{+}^{\otimes 3} =\\
    = p_2 p_3 \ket{F_1}\bra{F_1} 
    + (1-p_2) p_3 \ket{F_2}\bra{F_2} 
    + p_2 (1-p_3) \ket{F_3}\bra{F_3} +\\
    + (1-p_2) (1-p_3) \ket{F_4}\bra{F_4}\,.
\end{multline}

In~\cite{Multipartite_en_Salem_2024} this framework was used to study states of $3$ and $4$ qubits, associated to non-uniform hypergraphs.
Their analysis considered different entanglement measure, specifically the negativity and concurrence (for bipartite entanglement), and GMN (for multipartite entanglement) (see section~\ref{sec:ent_measures}).
Notably the authors showed that these quantities present a non-monotonic dependence on the gate success probabilities $p_{|e|}$.
Specifically the amount of entanglement as measured by these monotones can decrease and then increase again by lowering $p_{|e|}$ for certain edge cardinalities $|e|$.
This specific phenomenon is also called entanglement sudden death and birth~\cite{lopez2008sudden}.

In~\cite{Randomized_hype_Salem_2025} the authors extend the study to several configurations of $4$-qubit hypergraph states, extending the entanglement witnesses of Eq.~\eqref{eq:proj_W} for the case of mixed states, and applying them successfully to specific classes of hypergraphs.

Notably, this kind of randomization breaks LU-equivalence~\cite{Randomized_hype_Salem_2025}: if two pure hypergraph states $\ket{H}$ and $\ket{H'}$ are LU-equivalent, the corresponding randomized mixed states $\rho_P^H$ and $\rho_P^{H'}$ may not be, due to differing mixtures over subgraph supports. 
This implies that RHS can distinguish entanglement classes not resolved by LU transformations of their pure counterparts.

A slightly different approach was used in~\cite{Entanglement_of_Zhou_2022} where the authors define RHS directly as ensembles of pure hypergraph states.
Randomness is introduced by including each $k$-edge with a fixed probability $p = \frac{1}{2}$.
This corresponds to the previous case if we impose $p_{|e|} = \frac{1}{2}$ for all $|e|$.

To quantify entanglement, they focus on the subsystem purity and using tensor network techniques, they compute the average purity over the ensemble and show that both the case of $2$-edges (corresponding to standard graph states) and the case of $3$-edges satisfy the same scaling law for average entanglement.
Namely, for an equal bipartition with subsystems of size $n/2$ the average purity of the subsystem $A$ scales as $\Avg{P_A} \sim 2^{-n/2}$.
However, in the same case, the variance differs significantly:
\begin{align}
\Var{P_A}_3 &\sim 2^{-2n}, \\
\Var{P_A}_2 &\sim 2^{-n}.
\end{align}
Thus, states with $3$-edges exhibit typical entanglement behavior resembling Haar randomness~\cite{hamma2012ensembles}, differently from graph states.

\subsection{Nonclassical properties of hypergraph states}
\label{sec:nonclassicality}

Beyond their rich entanglement structure, hypergraph states exhibit many interesting nonclassical properties, such as strong violations of locality~\cite{brunner2014bell} and contextuality~\cite{budroni2022kochen}.
Both concepts constrain hidden variable models, but they are conceptually distinct.
The former works under the usual assumption found in (multipartite) Bell scenarios: that the outcome of one party is independent of the choice of measurement performed on a spatially separated subsystem.
The latter, as formalized by Kochen and Specker~\cite{kochen2011problem}, requires instead a stronger assumption, asserting that the outcome assigned to a measurement is independent of the measurement context (i.e. the set of compatible observables measured alongside it).
For our purposes, the relevance of KS-contextuality lies in the fact that it can be revealed even through nonlocal observables, provided they are mutually compatible. 
This allows one to detect contextuality using stabilizers of hypergraph states, which often involve nonlocal operators.

Similarly to the case of entanglement, for more than two parties we can also introduce the concept of \emph{genuine multipartite nonlocality} (GMNL)~\cite{svetlichny1987distinguishing}.
This concept extends Bell nonlocality to the multipartite setting by ruling out models where nonlocality is present within groups but not across them.
Specifically, a set of correlations is genuinely $n$-partite nonlocal if it cannot be decomposed into a convex combination of correlations that are local with respect to any bipartition.
Quantum states, such as GHZ, can violate Bell-type inequalities specifically designed to detect this form of nonlocality, thereby demonstrating that all $n$ parties are simultaneously involved in a nonlocal correlation.

\subsubsection{Stabilizer Formalism and Nonlocal Correlations}

Nonlocal correlations can be derived directly from the stabilizer structure~\cite{Entanglement_an_Guhne_2014, Extreme_Violati_Gachec_2016}.

For standard graph states a general recipe is known~\cite{guhne2005bell} for generating Bell inequalities tailored to an arbitrary graph state $\ket{G}$ in terms of their stabilizers.
Specifically, calling $S(G) = \{K_i\}_{i=1,\ldots,2^n}$ the set of the $2^n$ operators forming the stabilizer group for $G$, we can simply define a Bell operator as
\begin{equation}
    B(G) = \sum_{i=1}^{2^n} K_i\,.
\end{equation}
Since every $K_i \in S(G)$ is a product of local operators in the vertices of $G$, this is a well defined Bell-like inequality.
The quantum bound is given by $\expval{B(G)} \le b_Q(G) = 2^n$ while the classical bound $b_C(G)$ depends on the specific graph $G$.
Nonetheless it can be proved~\cite{guhne2005bell} that a gap always exists for any $G$, that is $b_Q(G) > b_C(G)$.

Unfortunately, this construction cannot be directly generalized to hypergraph states, due to the intrinsic nonlocality of the stabilizer operators in this case.

A preliminary study of nonclassicality properties of hypergraph states was done in~\cite{Entanglement_an_Guhne_2014}.
Focusing on $k$-uniform complete hypergraphs, the authors showed that using the stabilizer formalism for hypergraph states one can derive both contextuality and Bell-type inequalities.
The basic idea is to generalize the Mermin inequality~\cite{mermin1990extreme} for a complete $3$ qubit graph state $\ket{\Comp_3}$.
In this case the inequality can be written in terms of its stabilizers $K_1 = X_1Z_2Z_3, K_2 =Z_1X_2Z_3, K_3 = Z_1Z_2X_3$ as:
\begin{multline}
    \sum_i^{3} \Avg{K_i} + \Avg{K_1K_2K_3} = \Avg{X_1Z_2Z_3} + \Avg{Z_1X_2Z_3}+ \\
    +\Avg{Z_1Z_2X_3} -\Avg{X_1X_2X_3} \le 2
    \label{eq:mermin_k3}
\end{multline}
this quantity is classically bounded by $2$, and the bound is reached for a specific local $\pm 1$ assignment to the $X_i$ and $Z_i$ observables, but can reach the value $4$ when evaluated on $\ket{\Comp_3}$.
To generalize this to Mermin-like inequalities for hypergraph states we need the following condition to be satisfied by their stabilizers:
\begin{equation}
    \sum_i^n K_i + \prod_i^n K_i = \sum_i^n X_i \prod_{e \in E(i)} \CZ_{e\setminus \{i\}} - \prod_i^n X_i\,.
    \label{eq:mermin_hg}
\end{equation}
If~\eqref{eq:mermin_hg} is true, we can write an inequality in the form of~\eqref{eq:mermin_k3}.
Notice that the inequality is now defined in terms of the observables $\CZ_{e\setminus \{i\}}$, which, despite being compatible, are in general nonlocal, which makes the violation of such an inequality a suitable test for contextuality.
It can be shown that the condition~\eqref{eq:mermin_hg}, for a $k$-uniform complete hypergraph of $n$ qubits, is satisfied when \emph{i)} $\binom{n}{k}$ is odd and \emph{ii)} $\binom{n - l}{k - l}$ is even for any integer $1 \le l < k$.

Remarkably, these conditions can be easily visualized on the Pascal's triangle mod $2$, which approximates the fractal structure of the Sierpiński triangle, a pattern that was also noticed in~\cite{Local_Pauli_sta_Lyons_2017} where it was shown that it corresponds to the existence of a stabilizer of the form $-X_1\cdots X_n$ for the state (see also section~\ref{sec:local_stab}).

As mentioned, these inequalities generally serve as tests of contextuality; nevertheless, the same stabilizer-based approach can also be employed to derive nonlocality tests. 
This was already observed in~\cite{Entanglement_an_Guhne_2014} and was further analyzed in depth in~\cite{Extreme_Violati_Gachec_2016}.
To illustrate the idea consider the $3$ qubit hypergraph state defined by a single $3$-edge $\ket{\Comp_3^3} = \CZ_{(1,2,3)} \ket{+}^{\otimes 3}$.
Its stabilizer generators, as defined in~\eqref{eq:hg_stab}, are:
\begin{align}
    K_1 &= X_1 \CZ_{(2,3)}, \\
    K_2 &= X_2 \CZ_{(1,3)}, \\
    K_3 &= X_3 \CZ_{(1,2)}.
    \label{eq:stab_123}
\end{align}
Denote by $p(a_1 a_2 a_3 \mid P_1 P_2 P_3)$ the probability of measuring $a_1,a_2,a_3$ with $a_i = \pm 1$ when performing Pauli measurements $P_1, P_2, P_3$ on the qubit 1,2, and 3, respectively.
Since we can rewrite $K_1$ as
\begin{equation}
    K_1 = X_1 \otimes (\ket{00}\bra{00} + \ket{01}\bra{01} + \ket{10}\bra{10} - \ket{11}\bra{11})
\end{equation}
and knowing that $K_1 \ket{\Comp_3^3} = \ket{\Comp_3^3}$, we have that the conditional correlation $p(+-- \mid XZZ) = 0$. 
This is because the projector $P_{+11} = \ket{+11}\bra{+11}$ is necessarily orthogonal to the state $\ket{\Comp_3^3}$, since $P_{+11} K_1 = -P_{+11}$ hence we have $P_{+11} \ket{\Comp_3^3} = P_{+11} K_1 \ket{\Comp_3^3} = - P_{+11} \ket{\Comp_3^3}$.
By symmetry, one obtains the full set of forbidden outcomes $p(+-- \mid XZZ) = p(-+- \mid ZXZ) = p(--+ \mid ZZX) = 0$.
From these local constraints, a Hardy-type contradiction~\cite{hardy1993nonlocality} can be constructed.
Similarly, considering the projectors $P_{-00},P_{-01},P_{-10}$ we can conclude that $P(-++|XZZ) = P(-+-|XZZ) = P(--+|XZZ) = 0$, and the same is true if we permute the parties.

Assume a local hidden variable (LHV) model satisfying the above zero-probability conditions. Then it must also obey:
\begin{multline}
    p(+-- \mid XXX) + p(-+- \mid XXX) +\\+ p(--+ \mid XXX) = 0\,.
    \label{eq:mermin_hardytype}
\end{multline}

To see this, recall that in a LHV model we can assign a joint distribution for all the outcomes for each settings $X$ and $Z$ for all the parties.
Let us call $x_i$ and $z_i$ the $\pm1$ outcomes for party $i$ and measurement $X$ and $Z$ respectively.
Now consider the first term in \eqref{eq:mermin_hardytype}, in this case we have $x_1=+1$ and $x_2=x_3=-1$. 
Since $x_1=+1$ then, by the constraints we have on the forbidden outcomes, at least one between $z_2,z_3$ needs to be $+1$, but this is not possible since $x_2 = x_3 = -1$.
The other two terms in~\eqref{eq:mermin_hardytype} are also zero by symmetry.

However, quantum mechanics predicts 
\begin{equation}
p(+-- \mid XXX) = \frac{1}{16},    
\end{equation} 
resulting in a contradiction that can be used as a demonstration of the nonlocality of the state.
This Hardy-type construction can be promoted to a Bell inequality. 
Defining a Bell expression as:
\begin{multline}
    \Avg{B_3} = \left(p(+-- \mid XZZ) + p(-+- \mid XZZ) + \right.\\
    \left. +p(--+ \mid XZZ) +p(-++ \mid XZZ) + \text{sym.} \right) \\- \left( p(+-- \mid XXX) + \text{sym.} \right),
    \label{eq:h3_bell}
\end{multline}
we have that classically $\Avg{B_3} \geq 0$, but we find that quantum theory violates this inequality with $\Avg{B_3} = -3/16$.

The construction generalizes naturally to $n$-qubit hypergraph states with a single $n$-edge of maximum cardinality $\ket{\Comp_n^n}$.
In these states, the stabilizer formalism still yields perfect correlations for certain combinations of $X$ and $Z$ measurements. 
For instance, any outcome where a single party measures $a_{i_1} = +1$ and two others measure $a_{i_2} = a_{i_3} = -1$ is forbidden.
Quantum theory, however, predicts:
\begin{equation} 
    p(a_{i_1} = +1, a_{i_2} = a_{i_3} = -1 \ldots \mid P_1 \ldots P_n) =  \frac{1}{2^{n-2}},
\end{equation}
leading to a violation of the corresponding Bell inequality.
By extending this Bell expression with additional terms, one can even arrive at a generalized Svetlichny inequality~\cite{svetlichny1987distinguishing}, allowing to detect GMNL.

A further generalization of \eqref{eq:h3_bell} can be devised for $3$-uniform fully connected hypergraph states of $n$ qubits.
This can be done by noticing that $\Avg{X_1\cdots X_m Z_{m+1}\cdots Z_n}$ only depends on $m$, and specifically it is either $-1/2$ or $+1/2$ depending on whether $m = 0 \mod 4$ or $m = 2 \mod 4$ respectively.
For $\Avg{X_1\cdots X_n}$ instead we have either $0$ or $1$ for $n = 0 \mod 4$ or $n = 2 \mod 4$ respectively, while $\Avg{Z_1\cdots Z_n} = 0$ for any $n$.
This suggests the construction of the following Bell expression:
\begin{multline}
    B_n = -  Z_1\cdots Z_n +\\
    -\sum_{i=0}^{\floor{n/2}} (-1)^{i}\left(X_1\cdots X_{2i} Z_{2i+1}\cdots Z_n + \text{sym.} \right),
    \label{eq:h3_n_bell}
\end{multline}
which is bounded by 
\begin{equation}
    \Avg{B_n} \le 2^{\floor{n/2}}
    \label{eq:h3_n_bell_ineq}
\end{equation}
classically, but can reach the value of $\Avg{B_n} = 2^{n-2} - 1/2$ with quantum states.
The inequality defined above, gives an exponential violation (in the number of qubits) a property which is retained also for $4$-uniform fully connected hypergraph states.

Notably, for the $n$ qubit, fully connected $k$-uniform hypergraph states (with $k=3$ or $k=4$), one continues to violate the inequality~\eqref{eq:h3_n_bell_ineq} exponentially even when one or more qubits are traced out.

In~\cite{Symmetric_hyper_Noller_2023}, the authors refine and extend the derivation of these violations by rewriting~\eqref{eq:h3_n_bell} as:
\begin{equation}
    B_n^P = \frac{1}{2} \left( (P + iZ)^{\otimes n} + (P - iZ)^{\otimes n} \right),
    \label{eq:mermin_p_iz}
\end{equation}
where $P = X$ or $Y$ depending on the local stabilizer of the hypergraph state.
Let us denote with $\sqrt{P}^{\pm}$ the square root of the Pauli operator $P$ with eigenvalue $1$ and $\pm i$.
Consider now a $\vec k$-uniform hypergraph state $\ket{H^{\vec k}_n}$ stabilized by $P^{\otimes n}$. In this case we can simplify the computation of $\Avg{B_n^P}$.

By applying $(\sqrt{P}^-)^{\otimes n}$ to $B_n^P$, we obtain:
\begin{multline}
    B_n = (\sqrt{P}^-)^{\otimes n} B_n^P (\sqrt{P}^-)^{\otimes n} =\\
    =\frac{1}{2} \left( (\eye + iZ)^{\otimes n} + (\eye - iZ)^{\otimes n} \right).
    \label{eq:mermin_i_iz}
\end{multline}
This expression now only involves $Z$ operators, but it has the same bound of the original $B_n^P$ when evaluated on the transformed state $(\sqrt{P}^+)^{\otimes n} \ket{H^{\vec k}_n} = \ket{\tilde H^{\vec k}_n}$.

Specifically, by using this trick together with the decomposition of $\vec k$-uniform hypergraph states in \eqref{eq:GHZ_odd_decomposition}, one can find that the quantum expectation value is $\Avg{B_n^P} = 2^{n-2}$, while for any local hidden variable model we have $\Avg{B_n^P}\leq \sqrt{2}^n$.

Using this method in~\cite{Symmetric_hyper_Noller_2023} the authors proved the validity of this bound for a large class of $\vec k$-uniform states.
Specifically the result works for states $\ket{H^{\vec k}_n}$ such that $\vec k = (3,\ldots,2^m+1)$ or $\vec k = (2,3,\ldots,2^m+1)$ for some $m$, and with either $n=0\mod 2(2^m-1)$ or  $n=(2^m-1) \mod 2(2^m-1)$.
An example is the state $\ket{H_n^3}$ with $n = 0\mod 4$ or $n = 2\mod 4$, for which the authors also analyze the robustness of the violation under particle loss, obtaining that the state remains entangled up until the loss of $\floor{N-4}/2$ particles.

\begingroup
\setlength{\tabcolsep}{7pt} 
\renewcommand{\arraystretch}{1.5} 

\begin{table*}[htp]
\centering
\begin{tabular}{llm{7cm}m{3.5cm}}
\toprule
\textbf{Operation} & \textbf{Symbol} & \textbf{Graphical description} & \textbf{Example} \\
\midrule
Pauli-$Z$ & $Z_i$ & Introduces the edge $\{i\}$ of cardinality one. & \includegraphics[height=4em]{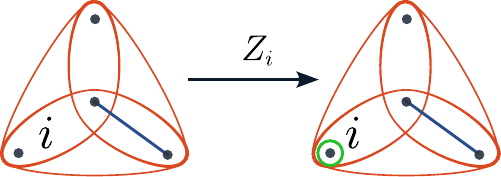} \\
Pauli-$X$ & $X_i$ & Toggles the presence of all hyperedges in the adjacency set $A(i) = \{e \setminus \{i\} : e \in E(i)\}$. & \includegraphics[height=4em]{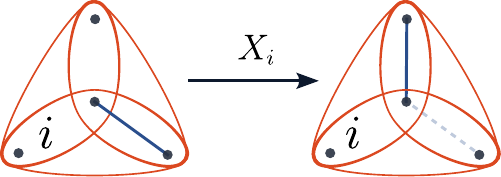} \\
Control-$Z$ & $\CZ_e$ & Fundamental building block: adds/removes an hyperedge $e$. & \includegraphics[height=4em]{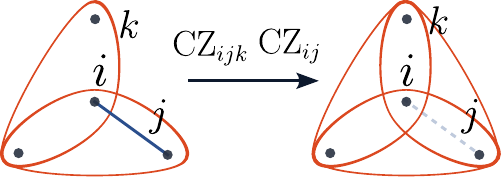} \\
Control-NOT & $\CCNOT{k}_{S,T}$ & Removes or introduces new edges defined by the union of the controls $S$ and the adjacent edges to the target $t$: $E \mapsto E \sdiff \{e \cup S : e \in A(t)\}$. & \includegraphics[height=4em]{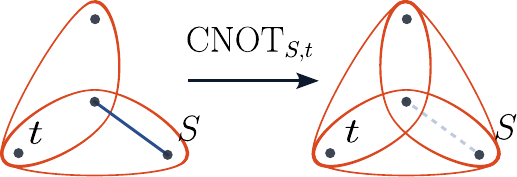} \\
EPC & $\sqrt{X}^\pm\prod_{e\in E}\sqrt{\CZ_e}^\pm$ & \textbf{Edge-Pair Complementation} (EPC): toggles hyperedges formed by the union $e \cup f$ of pairs of edges $e, f\in A(i)$. & \includegraphics[height=4em]{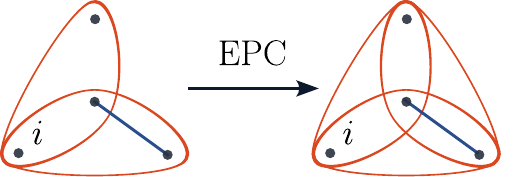} \\
$Z$ Measurement & $\Pi_i^z$ & Effectively removes the vertex and all incident hyperedges. & \includegraphics[height=4em]{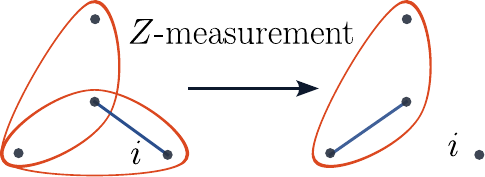} \\
$X$ Measurement & $\Pi_i^x$ & In general does not admit a graphical description (see section~\ref{sec:MBQC_hg}). & \\
\bottomrule
\end{tabular}
\caption{Common operations for qubit hypergraph states and their descriptions in terms of their hypergraph representation.}
\label{tab:qb_ops}
\end{table*}

\endgroup

\section{Hypergraph states and computation}
\label{sec:computation}

\subsection{Measurement-Based Quantum Computation (MBQC)}
\label{sec:MBQC_intro}

Measurement-based quantum computation (MBQC)~\cite{raussendorf2001one} offers a paradigm shift from the traditional circuit model by enabling universal quantum computation through a sequence of adaptive single-qubit measurements on a highly entangled many-body resource state. In this model, the quantum input is prepared into a particular resource state, and computation is driven forward by measurements, with classical feed-forward needed to perform adaptively the next measurement in specific bases that depend on the previous outcomes.

The canonical resource in MBQC has historically been the cluster state, which falls in the class of graph states.
Recent advances~\cite{Quantum_computa_Takeuc_2019, Changing_the_ci_Gachec_2019}, have demonstrated that hypergraph states can extend the power and flexibility of MBQC in some tasks.
Notably, while requiring a larger set of gates for the generation of the initial state (which includes at least the CCZ gate), they require only measurements on Pauli bases, and they can have advantages in term of gate parallelization.

In standard MBQC, the computation is performed via single-qubit measurements on a fixed entangled state, typically a graph state such as a 2D cluster state~\cite{raussendorf2003measurement}. 
The computation proceeds by choosing measurement bases adaptively based on earlier measurement outcomes, which allows the simulation of any unitary gate and perform universal computation.

The key resources that characterize a specific MBQC protocol are the gates required for the state preparation and the basis in which the adaptive measurements need to be performed.
We can use the Clifford hierarchy $\mathcal{C}_k$, to classify these resources, which can be defined recursively as:
\begin{align}
\Cliff_1 &= \Pauli_n, \\
\Cliff_{k+1} &= \left\{ U \mid \forall P \in \Cliff_1, UPU^\dagger \in \Cliff_k \right\}.
\end{align}
where $\Pauli_n$ is the set of of Pauli operators on $n$ qubits.

Particularly important is the second level of this hierarchy, $\Cliff_2$, which is the set of operations that preserve the Pauli group $\Pauli_n$.
Differently from $\Cliff_k$ with $k \ge 3$, $\Cliff_2$ forms a group called the Clifford group, which we will also denote by just $\Cliff$.
This correspond to the group generated by:
\begin{itemize}
    \item The Pauli group $\Pauli_n$.
    \item The Hadamard $H$ and phase gate $S$ defined in~\eqref{eq:H_S}.
    \item The $\CNOT$ gate (or alternatively the $\CZ$ gate) on two qubits.
\end{itemize}

It is well known that the Clifford group $\Cliff_2$ enables efficient classical simulation under Pauli measurements~\cite{aaronson2004improved}, while universal quantum computation requires gates from at least $\Cliff_3$.

Adaptive measurements in this scheme are a necessity to correct the by-product operations on the subsystems due to earlier measurements.
In the standard approach to MBQC these by-products are represented by the Pauli $X$ and $Z$ operators.

\subsubsection{MBQC Protocols with Hypergraph States}
\label{sec:MBQC_hg}

\begin{figure*}
    \begin{subfigure}{0.3\textwidth}
        \includegraphics[width=\textwidth]{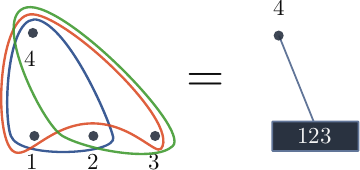}        
    \caption{Basic building block of the hypergraph state used in the MBQC protocol.}
    \label{fig:mbqc_bb}
    \end{subfigure}
    \hspace{0.1\textwidth}
    \begin{subfigure}{0.3\textwidth}
        \includegraphics[width=\textwidth]{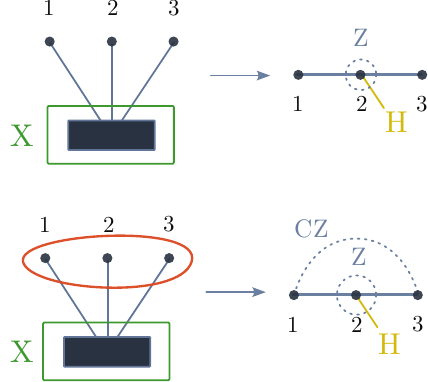}
    \caption{Graphical representation of the action of a $X$ measurement.}
    \label{fig:mbqc_X}
    \end{subfigure}
    \caption{Representation of the basic building block of the hypergraph state needed for the protocol, and effect of the Pauli $X$ measurement on it~\cite{Changing_the_ci_Gachec_2019}.
    To simplify the notation, the box stands for three nodes, and a single line from the box to another node means that the latter shares a $3$-edge with any couple of the other three.
    The action of the Pauli $X$ is represented by the solid green box, Hadamard correction are indicated by the yellow line while  dashed lines and circle represents the $Z$ and the $\CZ$ byproduct.}
    \label{fig:mbqc_basics}
\end{figure*}

The protocol proposed in~\cite{Changing_the_ci_Gachec_2019} uses $3$-uniform hypergraph states generated by CCZ (or $\CCZ{3}$ in our notation) gates, which are in $\Cliff_3$.
The specific structure of the states is shown in Fig.~\ref{fig:mbqc_basics}.
However, this approach has the advantage to be able to perform deterministic computation using only Pauli $\{X,Y,Z\}$ measurements, without the need for measurement in different bases.
Moreover, it also enables the parallel execution of all logical CCZ and SWAP gates. 

However, in this case the resulting by-product set is given by $\{ \CZ, X, Z \}$, which includes also the $\CZ$ gates in constrast with standard MBQC on graph states.

We will now give a coarse description of how the scheme works, starting from the graphical rules governing the Pauli measurement when applied on the hypergraph state (see table~\ref{tab:qb_ops}).

A measurement on the $Z$ basis, simply detaches the node from the graph, just like in the usual MBQC protocols.
Conversely, $X$ measurements on hypergraph states generally lead outside the space of hypergraph states. 
However, under certain structural conditions, they can deterministically produce states that are still locally equivalent to hypergraph states.
This can be derived using the graphical rule obtained from generalized local complementation described in~\ref{sec:graphical_ops_lu}.
Regarding this specific protocol, we are interested in the effect of this measurement on three of the $6$ qubits that form the $3$-uniform hypergraph depicted in Fig.~\ref{fig:mbqc_X}.
The effect of the measurement is to produce $2$-edges between the three remaining qubits, with additional byproduct given by a Hadamard $H$ or a $\CZ$.

\begin{figure}[b]
    \centering
    \includegraphics[width=0.5\linewidth]{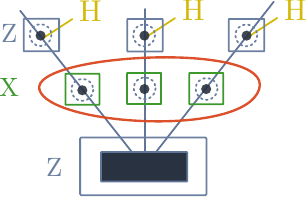}
    \caption{Graphical representation of the procedure for the application of a nearest neighboroud $\ccz$, using Pauli $X$ measurements~\cite{Changing_the_ci_Gachec_2019}.
    As in Fig.~\ref{fig:mbqc_basics} the Pauli $X$ and $Z$ are denoted by the solid green and blue box respectively, dashed lines and circle are used for the $Z$ and the $\CZ$ byproduct and Hadamard correction are indicated by the yellow line.}
    \label{fig:teledge}
\end{figure}

To show universality we can choose the set $\{\ccz, H \}$, which is sufficient for universal quantum computation.
To implement a logical $\ccz$ gate on arbitrary qubits, one begins with applying the $\ccz$ gate on nearest-neighbor.
Starting from an initial $3$-uniform resource we first generate $2$-edges appropriately by applying Pauli $X$, using the aforementioned rule.
Then we perform a teleportation step, also via Pauli $X$ measurements, of a given $3$-edge $e$ to the three other qubits connected by $2$-edges to each of the nodes in $e$ (see Fig.~\ref{fig:teledge}).
After that we only need to implement SWAP gates to be able to apply the $\ccz$ operator to an arbitrary set of nodes.
This can then be realized by further reducing the hypergraph to an hexagonal lattice structure with only $2$-edges, by measuring auxiliary nodes in the Pauli $X$ basis, and removing nodes appropriately via $Z$ measurements.  
This structure is in general suitable for implementing any Clifford operation deterministically, up to $X$ and $Z$ by-products.

All logical CCZ gates, even those spanning distant qubits, can be implemented in parallel. 
This contrasts with the cluster-state MBQC where only Clifford gates benefit from such parallelism.
However, since the by-product group $\{\CZ, X, Z\}$ is not closed under conjugation by the Hadamard gate, all CZ by-products must be corrected prior to the application of any global Hadamard transformation.
The logical depth of the protocol can then be associated with the number of Hadamard layers. 
Since CCZ operations can be parallelized, the depth increases only when Hadamard layers are applied.

This gives this protocol advantage over other computational schemes for certain tasks.
For example, considering the $\CCZ{k}$ gate acting on $k$ qubits, the resulting circuit achieves logarithmic depth with respect to $k$, and is linear in the number of $H$, $\CZ$ and ancillae required, while other implementations for the same operation usually require either exponentially more $\CZ$ operations and ancillae, or a higher depth. 

A similar protocol was also proposed in~\cite{Quantum_computa_Takeuc_2019}, which has the advantage of using only measurements in the $X$ and $Z$ basis.
However, it uses a different hypergraph state as its universal resource, presenting both $2$ and $3$-edges, which needs both $\CZ$ and $\ccz$ gates to be generated, and does not allow for the parallelization of the $\ccz$ and SWAP gates.

\subsection{Non-stabilizerness}
\label{sec:nonstab}

The Gottesman-Knill theorem~\cite{aaronson2004improved} provides an example of classical simulability in quantum systems by demonstrating that stabilizer states, despite their entanglement, can be efficiently simulated on a classical computer~\cite{PhysRevA.70.052328}. 
As a result, achieving universal fault-tolerant quantum computation necessitates the use of non-stabilizer states—those that lie beyond the stabilizer formalism \cite{bravyi2005universal, howard2017application}. 
These non-stabilizer states embody a key computational resource known as \emph{magic} \cite{bravyi2005universal}, or non-stabilizerness, which is essential for realizing quantum computational advantage~\cite{gottesman1997stabilizer}.

Formally stabilizer states can be defined as those states that can be generated from the computational basis by means of Clifford operations:
\begin{definition}[Stabilizer states]
    A pure state $\ket{\psi}$ on $n$ qubits is called a \emph{stabilizer state} iff it can be written as
    \begin{equation}
        \ket{\psi} = C\ket{0}^{\otimes n}\,,
    \end{equation}
    where $C \in \Cliff$ is in the Clifford group.
\end{definition}

Alternatively one can define the set of pure stabilizer states on $n$ qubits as those states that are eigenstates with eigenvalue $+1$ of all the operators in an Abelian subgroup $S \subset \Pauli_n$ of the Pauli group.
From this characterization, and the discussion in section~\ref{sec:hg_stabilizers}, we immediately see that all graph states are also stabilizers states.

We can also define the set $\STAB$ of all the stabilizer states containing also their convex mixtures:
\begin{equation}
    \STAB = \left\{\sum_ip_i\ket{\psi_i} : \sum_i p_i = 1 \,,\, \{\ket{\psi_i}\}\,\text{are stabilizers}\right\}\,.
\end{equation}

The requirement of non-stabilizerness comes from the fact that the stabilizer formalism, with Clifford operations and stabilizer states, constitutes a classically tractable subtheory of quantum mechanics~\cite{bravyi2005universal}. 
To avoid this limitation, one must either include non-Clifford quantum gates, i.e. those that cannot be generated from Clifford operations, or introduce specially prepared resource states, known as magic states.

When employed as ancillae, these magic states allow to attain universal quantum computation, via procedures such as magic state injection and distillation~\cite{Souza2011,10.1007/978-3-642-10698-9_3}.
Hypergraph states, in contrast to simple graph states, are a clear example of magic states, as we have seen in the previous section, where we show that they are a crucial resource for universal quantum computation in MBQC protocols without non-Clifford operation.

The resource-theoretic approach to quantum computation has led to formal definitions and quantitative characterizations of magic.
Several quantities have been proposed to capture different aspects of non-stabilizerness, including the \emph{robustness of magic}~\cite{PhysRevLett.118.090501, Heinrich2019robustnessofmagic, Hamaguchi2024handbookquantifying}, \emph{mana}~\cite{Veitch_2014}, and \emph{stabilizer entropies}~\cite{PhysRevLett.128.050402}.

In the following we will briefly review some of the monotones that were used in the literature to describe non-stabilizerness specifically in the case of hypergraph states.

\paragraph*{Relative entropy of magic.}
We start by considering two measures of non-stabilizerness based on generalized Petzl-Reny\'i relative entropies~\cite{petz1986quasi}, which for quantum states are defined by:
\begin{equation}
    D_\alpha(\rho \|\sigma) = \frac{1}{1-\alpha}\log\Tr\left(\rho^\alpha \sigma^{1-\alpha}\right)\,.
\end{equation}
When $\mathrm{supp}(\rho) \subseteq\mathrm{supp}(\sigma)$, where $\mathrm{supp}$ is the support of the state, and $\infty$ otherwise.
For $\alpha \to 1$ we recover the usual quantum relative entropy, $D(\rho \| \sigma) = \Tr \rho (\log\rho - \log \sigma)$, while the case $\alpha=0$ and $\alpha \to \infty$ are called min- and max-relative entropy respectively:
\begin{align}
     &D_{\infty}(\rho \|\sigma) = D_{\max}(\rho \|\sigma) = \log \min\{\lambda : \rho \le \lambda\sigma\}\,,\\
     &D_{0}(\rho \|\sigma) = D_{\min}(\rho \|\sigma) = -\log \Tr(\Pi_\rho\sigma)\,,
\end{align}
where $\Pi_\rho$ is the projector onto the support of $\rho$.
Intuitively $D_{\min}$ is the most ``relaxed'' notion of distance among the Petzl-Reny\'i entropies, and considers the difference between the states by comparing how much one state is supported on the space of the other.
Conversely $D_{\max}$ captures the ``worst case'' distance in the direction where the states are most different.

The \emph{Min-relative entropy of magic} captures the minimal distinguishability between a quantum state and the set of stabilizer states, in terms of the min-relative entropy.
It is defined for a quantum state $\rho$ as:
\begin{equation}
\Dm_{\min}(\rho) = \min_{\sigma \in \STAB} D_{\min}(\rho \| \sigma),
\end{equation}
where $\STAB$ is the set of stabilizer states.
For pure states $\ket{\psi}$, this simplifies to
\begin{equation}
\Dm_{\min}(\psi) = -\log \max_{\phi \in \STAB} |\langle\psi | \phi\rangle|^2\,.
\end{equation}
which indicates how well a stabilizer state can approximate the given state in terms of overlap.

Conversely the \emph{Max-entropy of magic} considers the worst case scenario for distinguishing a state from stabilizer states. 
It is given by:
\begin{equation}
\Dm_{\max}(\rho) = \min_{\sigma \in \STAB} \log \min\{\lambda \mid \rho \leq \lambda \sigma\}\,.
\end{equation}

Its meaning is perhaps clearer if we notice that is equivalent to the logarithmic version of another monotone called \emph{generalized robustness} $R_g$
\begin{equation}
\Dm_{\max}(\rho) = \log(1 + R_g(\rho)),
\end{equation}
which characterizes how much a state must be ``diluted'' to fall within the stabilizer set, and it is defined as

\begin{align}
&R_g(\rho) =\min\{\alpha : \rho = (1 - \alpha) \sigma_\STAB + \alpha \sigma_\States\} \\
&\text{s.t.}\quad \sigma_\STAB \in \STAB, \,\sigma_\States \in \States
\end{align}
where $\States$ represents the set of all states.

Notably, for any quantum states $\rho$, these monotones satisfy the condition:
\begin{equation}
\Dm_{\min}(\rho) \leq \Dm_{\max}(\rho) \,.
\end{equation}

\paragraph*{Stabilizer Rényi Entropy.}
Another important measure of non-stabilizerness is given by the \emph{Stabilizer Rényi-$\alpha$ Entropy (SRE)}\cite{leone2022stabilizer}, which is defined for a pure quantum state $\ket{\Psi}$ as follows:
\begin{align}
    M_\alpha(\ket{\Psi}) = \frac{1}{1 - \alpha} \log \left( \sum_{P \in \mathcal{P}_n} \left( 2^{-n} \mathrm{Tr}(P \ket{\Psi}\bra{\Psi})^2 \right)^\alpha \right) - n,
\end{align}
where $\Pauli_n$ is the $n$-qubit Pauli group containing only $+1$ phases.
This can be seen as the Rényi entropy of the distribution $p_i = 2^{-n}\Tr(P_i \ket{\Psi}\bra{\Psi})^2$ for $P_i \in \Pauli_n$, which is a properly defined probability for pure state since we can expand its purity $\Tr{\rho^2}$ as:
\begin{equation}
    \Tr{\rho^2} = \sum_{P_i \in \Pauli_n} 2^{-n}\Tr(P_i \ket{\Psi}\bra{\Psi})^2 = 1.
\end{equation}
The fact that this quantity is zero for all stabilizer states, follows from the fact that for any $\ket{\Psi} \in \STAB$ we have $\expval{P}_{\Psi} = \pm 1$ for exactly $2^n$ elements of $\mathcal{P}_n$ and zero for all the others.
Moreover we have $M_\alpha(\Psi) > 0$ whenever $\ket{\Psi} \notin \STAB$.
Therefore $M_\alpha(\Psi)$ is a faithful measure that quantifies the deviation of $\Psi$ from the set of stabilizer states.

An equivalent form can be expressed using the \emph{Pauli-Liouville moment}:
\begin{align}
    m_\alpha(\ket{\Psi}) = 2^{-n} \sum_{P \in \mathcal{P}_n} \left( \mathrm{Tr}(P \ket{\Psi}\bra{\Psi})^2 \right)^\alpha,
\end{align}
so that:
\begin{align}
    M_\alpha(\ket{\Psi}) = \frac{1}{1 - \alpha} \log m_\alpha(\ket{\Psi}).
\end{align}

\subsubsection{Magic in Hypergraph states}

The non-stabilizerness of hypergraph states is closely connected to their entanglement structure.
In section~\ref{sec:nonstab} we have showed that an important difference between graph and hypergraph states is the nonlocal and non-Clifford nature of the generalized stabilizer operators for the latter,
while the former are stabilized by local Paulis.
This form of non-stabilizerness is also called \emph{nonlocal magic}~\cite{qian2025quantum}, to distinguish it from the one that can be traced back to non-stabilizerness of the individual subsystems. 

A first study of the non-stabilizer properties of hypergraph states was performed in~\cite{Many_Body_Quant_Liu_Z_2022}.
Here the authors use the fact that for REW states, hence for hypergraph states, whenever the function $f$ defining the state $\ket{H_f}$ as in~\eqref{eq:REW_definition} is a quadratic Boolean functions, $\ket{H_f}$ is in the set of stabilizer states (although the converse is not true in general).
Only higher degree functions can introduce non-stabilizerness.

To measure how far a Boolean function $f$ is from being quadratic, one can use the notion of \emph{non-quadraticity}:
\begin{equation}
\chi(f) = \min_{q \in \mathcal{Q}} w(f + q),
\end{equation}
where $\mathcal{Q}$ is the set of quadratic Boolean functions, and $w(f)$ denotes the Hamming weight of a Boolean function $f$, defined as the number of ones in the bitstring corresponding to the truth table of $f$.
Specifically the number $w(f+q)$ represents the number of collisions of $f$ and $q$ and it is also called the Hamming distance of between $f$ and $q$.
This distance influences the overlap between the corresponding hypergraph state $\ket{H_f}$ and the stabilizer states, yielding the following upper bound on the min-relative entropy of magic:
\begin{equation}
\Dm_{\min}(\ket{H_f}) \leq -2\log\left(1 - 2^{-n+1} \chi(f)\right).
\label{eq:bound_dmin_chi}
\end{equation}
Moreover, using a connection with classical coding theory via the Reed-Muller codes~\cite{abbe2020reed}, one can also prove a general upper bound on $\Dm_{\min}$ valid for any hypergraph state.
\begin{equation}
\chi(f) \leq 2^{n-1} - \frac{\sqrt{15}}{2} 2^{n/2},
\end{equation}
and using~\eqref{eq:bound_dmin_chi} one obtains:
\begin{equation}
\Dm_{\min}(\ket{H_f}) \leq n - \log 15 + o(1)\,,
\label{bound_dmin_hg}
\end{equation}
which corresponds to the scaling that we expect for the magic of a typical random state.

A more in depth analysis of the non-stabilizer properties of hypergraph states was performed in~\cite{Magic_of_quantu_Chen_2024}.
Here the authors derive an analytical formula for the Rényi entropy of magic $M_\alpha(\ket{H})$, using which they prove that one can bound the magic of the state from the the average degree $\bar{\Delta}(H)$.
Recall, that the degree of a vertex in a hypergraph $\Delta(v)$ is the cardinality of its neighboring set, i.e. the set of the vertices $u$ such that ${v, u} \subseteq e$ for some edge $e \in E$ of the hypergraph.
The average $\bar{\Delta}(H)$ is taken on all the vertices $v\in V$ of $H$.
Explicitly the bound can be written as:
\begin{align}
    M_\alpha(\ket{H}) \leq \frac{1}{\alpha - 1} \left[1 - \log \left(1 + \frac{1}{2(2^\alpha - 1)\bar{\Delta}(H)}\right) \right] n.
\end{align}
This implies that maximum magic cannot be reached for particular classes of hypergraph where the degree is bounded, like for instance hypergraph states on a lattice~\cite{miller2018latent}.

Using this quantity one can also study the non-stabilizerness of random hypergraph states (with uniformly sampled hyperedges), showing that random $k$-uniform hypergraph states ($k \geq 3$) typically have near-maximal magic:
\begin{align}
    \Avg{M_\alpha} \geq \frac{n - (k+2^{2\alpha -1})}{\alpha-1},
\end{align}
and their variance shows a ``concentration effect'' of magic, implying that most random hypergraph states are maximally magical, similarly to what happens with Haar random states~\cite{Many_Body_Quant_Liu_Z_2022, leone2022stabilizer}.

Focusing instead on the symmetric $k$-uniform hypergraphs with $n$ qubits, it is possible to find explicit results for the moments $m_2(\ket{\Comp_n^k})$ and $m_{1/2}(\ket{\Comp_n^k})$.
This in turn allows to derive some information on the behavior of $M_2(\ket{\Comp_n^k})$ and $M_{1/2}(\ket{\Comp_n^k})$.
One finds that for $k=3$, $M_2$ is bounded by a constant, while $M_{1/2}$ is linear in $n$.
The situation is very different for $k=n$ where $M_2$ decreases exponentially with $n$, while for large $n$, $M_{1/2}$ increases monotonically to $M_{1/2} \to 2\log_2 3$.
This also shows that these states may have large $M_{1/2}$ even when $M_2$ is small, a strikingly separation in behavior depending on $\alpha$.

\subsection{Error correction}

Quantum information stored in physical qubits is inevitably subject to noise arising from interactions with the environment, imperfect control, and decoherence processes.
Unfortunately, unlike classical information, quantum states cannot be cloned and measured directly without disturbance, which makes the protection of quantum information a fundamentally different task. 
Quantum error correction (QEC) protocols aim to do this by encoding logical qubits into carefully chosen subspaces of a larger Hilbert space such that the effects of certain kinds of errors can be detected and reversed without degrading the logical state itself.

A possible choice of code spaces for QEC is represented by graph states~\cite{schlingemann2001quantum,looi2008quantum}, these are stabilizer codes~\cite{cleve1997efficient}, where the graph structure naturally encodes the stabilizer generators.
Sharing a similar structure, hypergraph states were also proposed as a natural generalization of QEC with graph states~\cite{Quantum_Error_C_Balaku_2017}.
In this section we will see that they indeed allow to generate new interesting families of quantum codes, such as the symmetric hypergraph codes~\cite{Analysis_of_qua_Wagner_2018}.

One of the goals of the general framework of QEC is to determine if, for a given encoding, there exists a recovery operation that perfectly restores all states in the code subspace. 
This is given by the Knill--Laflamme condition, a necessary and sufficient criterion characterizing exactly when a code can correct a given set of errors. 
This condition provides an important foundation for the theory of error-correcting codes, and will serve as the starting point for the constructions discussed in the following sections.

Consider a Hilbert space of $n$ physical qubits $\Hil \cong (\CC^2)^{\otimes n}$ and let $\mathcal{C}\subset\Hil$ be a $K$-dimensional code subspace with projector $P$.
The noise on the system can be modeled by a completely positive trace-preserving (CPTP) map $\mathcal{E}$ with Kraus (error) operators $\{E_i\}_i$:
\begin{equation}
\mathcal{E}(\rho)=\sum_i E_i \rho E_i^\dagger,\qquad \sum_i E_i^\dagger E_i = I.
\label{eq:noise_chan}
\end{equation}
We say that $\mathcal{C}$ \emph{corrects} the error set $\{E_i\}_i$ if there exists a recovery channel $\mathcal{R}$ such that for all $\ket{\psi}\in\mathcal{C}$,
\begin{equation}
\mathcal{R}\circ \mathcal{E} \left(\ket{\psi}\bra{\psi}\right)\propto \ket{\psi}\bra{\psi}.
\end{equation}
The channel $\mathcal{R}$ allows us to abstract away from the actual implementation of the protocol, and focus on the relevant properties of the error correcting code.

\begin{theorem}[Knill--Laflamme (KL) condition]
\label{thm:KL}
A code $\mathcal{C}$ corrects the error set $\{E_i\}_i$ if and only if there exist complex numbers $\alpha_{ij}$ (depending only on $i,j$) such that
\begin{equation}
\label{eq:KL-matrix}
\bra{\phi} E_i^\dagger E_j \ket{\psi} = \alpha_{ij}\,\braket{\phi|\psi}
\quad\text{for all }\ket{\phi},\ket{\psi}\in\mathcal{C}.
\end{equation}
Equivalently, in operator form,
\begin{equation}
\label{eq:KL-projector}
P\,E_i^\dagger E_j\,P=\alpha_{ij}\,P\qquad\text{for all}\qquad i,j.
\end{equation}
\end{theorem}
The off-diagonal requirement states that if $\ket{\phi}\perp\ket{\psi}$, then $\bra{\phi}E_i^\dagger E_j\ket{\psi}=0$,
while the diagonal requirement is that the expression only depends on $i$ and $j$ but not on the particular codeword label.

If $\alpha_{ij}$ happens to be diagonal, i.e. $\alpha_{ij}=c_i\delta_{ij}$, then distinct correctable errors map the code to mutually orthogonal subspaces.
Notably, in the more general case, $\alpha_{ij}$ may have nonzero off-diagonal elements, meaning that different errors on the same codeword can generate overlapping states.
Such codes are called degenerate, and in principle are more efficient in terms of the dimension of the code space $\mathcal{C}$.

Notice that if the KL condition holds for the Pauli set $\{I_i,X_i,Y_i,Z_i\}$ on each qubit, it immediately follows that arbitrary single-qubit noise on that site is correctable. 
In this case it will be useful to introduce the notion of \emph{distance}.
For a Pauli operator $P$, define its weight $w(P)$ as the number of qubits on which it acts nontrivially.
The distance $d$ of a code is the largest integer such that the KL condition holds for all products $E_i^\dagger E_j$ of correctable errors with $w(E_i^\dagger E_j) < d$:
A code of distance $d$ corrects all errors on up to $t=\lfloor (d-1)/2\rfloor$ qubits, or equivalently all Pauli errors of weight at most $t$.

\subsubsection{Quantum error correction with hypergraph codes}

As mentioned, hypergraph states can be used successfully for constructing quantum error-correcting codes.
A possible way to do this is by using symmetric hypergraph states~\cite{Analysis_of_qua_Wagner_2018}.
Recall that a \emph{symmetric hypergraph state} on $n$ qubits is one that remains invariant under any permutation of the qubits (see section~\ref{sec:definitions}) and that their $\pm 1$ amplitudes in the computational basis can only depend on the Hamming weight of the basis states.
Hence they can be succinctly defined by only specifying the set of weights that acquire negative amplitudes $M = \{m_1,\ldots,m_k\}$, i.e. such a state has $-1$ amplitude for a given computational basis state $\ket{x} = \ket{x_1,\ldots,x_n}$ if and only if the Hamming weight $w(x)$ of $x$ is contained in $M$.
The symmetry properties will be useful to ensures that all qubits are treated on an equal footing, which simplifies the verification of the KL conditions.

Any symmetric hypergraph code $\mathcal{C}$ on $n$ qubits can be generated given a symmetric hypergraph $H=(V,E)$ and a subset of the nodes $I\subseteq\{1,\dots,n\}$.
Such a code is defined by $\mathcal{C} = \mathrm{span}\{\ket{H}, \ket{H_I}\}$ where $\ket{H_I} =  \prod_{i \in I} Z_i \ket{H}$, is an element of the hypergraph basis of $\ket{H}$ (see section~\ref{sec:hg_stabilizers}).

To analyze this class of codes it is useful to introduce the notion of \emph{$X$-difference hypergraph}.
Consider a hypergraph $H=(V,E)$ with vertex set $V=\{1,\dots,n\}$. 
For a fixed vertex $i\in V$, the $X$-difference hypergraph $D_i$ is defined by the reduced vertex set $V\setminus\{i\}$ and the set of edges $E_{D_i}$ given by
\begin{equation}
E_{D_i} = \left\{ e\setminus\{i\} : e \in E, i \in e \right\}.
\end{equation}
That is, $D_i$ is obtained by removing vertex $i$ from the hypergraph and, for every hyperedge containing $i$, adding the corresponding edge with $i$ deleted. 

The reason why $D_i$ is called $X$-difference hypergraph is because it can be written as a hypergraph symmetric difference.

The \emph{symmetric difference of two hypergraphs} $H_1=(V, E_1)$ and $H_2=(V,E_2)$, denoted as $H = H_1 \sdiff H_2$, is defined as the hypergraph $H = (V, E_1 \sdiff E_2)$, where $E_1 \sdiff E_2 = (E_1 \cup E_2)\setminus (E_1 \cap E_2)$ is the symmetric difference of the sets of their edges.

Similarly we can define the \emph{symmetric difference of two hypergraphs states} simply as the symmetric difference of their hypergraph $\ket{H_1}\sdiff\ket{H_2} = \ket{H_1 \sdiff H_2}$.

Applying the $X_i$ operator to $\ket{H}$ yields
\begin{equation}
(\ket{H}) \sdiff (X_i\ket{H}) = \ket{+}_i \otimes \ket{D_i},
\label{eq:xdiff_sd}
\end{equation}
up to a global phase.

Now, let $D$ be the $X$-difference hypergraph of the symmetric hypergraph $H$ used to defined the code $\mathcal{C}$ (we can safely drop the index $i$ in $D_i$since $H$ is symmetric).
One can show~\cite{Analysis_of_qua_Wagner_2018} that $D$, together with the cardinality of the set $I$, is sufficient to completely define $\mathcal{C}$.
Moreover, since $\ket{D}$ is itself symmetric, it can be described by a set of integers $m_1,\ldots,m_k \in \{1,\dots,n-1\}$ denoting the weights with negative amplitudes in the computational basis.
In this way, a unique tuple $\left(\{m_1,\ldots,m_k\},l\right)$ with $l=|I|$ can be associated with every symmetric hypergraph code $\mathcal{C}$.
Conversely, every such tuple defines a unique code (on a given number of qubits).
This establishes a one-to-one correspondence between codes and combinatorial data.

The error detectability conditions can then be expressed in terms of the \emph{balancedness} of $\ket{D}$.
A state $\ket{\phi}$ on $m$ qubits is called balanced if it contains equally many positive and negative coefficients in the computational basis representation, or equivalently if 
\begin{equation}
\bra{+}^{\otimes m}\! \ket{\phi} = 0.
\end{equation}

Notably, the scalar product of two hypergraph states $\ket{H}, \ket{H'}$ depends only on their symmetric difference as $\braket{H|H'} = \bra{+}^{\otimes n} \! (\ket{H} \sdiff \ket{H'})$, which means that they are orthogonal if and only if $\ket{H} \sdiff \ket{H'})$ is balanced.
It follows that if $\ket{D}$ is balanced, from~\eqref{eq:xdiff_sd}, the $X_i$ errors map $\ket{H}$ into a state orthogonal to the code space, making them detectable.
This reduces the problem of verifying the KL condition to checking the balancedness of $\ket{D}$.
Using this one can determine when these codes achieve a nontrivial distance.
From the KL condition, one can prove that if $l>1$, then $\mathcal{C}$ has distance at least $2$ if and only if all of the following conditions hold:
\begin{enumerate}
\item $\ket{D}$ is balanced,
\item $\big(\prod_{i=1}^l Z_i\big)\ket{D}$ is balanced,
\item $\big(\prod_{i=1}^{l-1} Z_i\big)\ket{D}$ is balanced.
\end{enumerate}

These balancedness conditions can also be rewritten in purely combinatorial terms involving binomial coefficients, leading to the following classification criterion:
\begin{align}
\sum_{m\in M} \binom{n}{m} &= 2^{n-1}, \\
\sum_{m\in M} \sum_{\substack{j\ \text{odd}\\ j\leq l,\, j\leq m}}
\binom{l}{j}\binom{n-l}{m-j} &= 2^{n-2}, \\
\sum_{m\in M} \sum_{\substack{j\ \text{odd}\\ j\leq l-1,\, j\leq m}}
\binom{l-1}{j}\binom{n-(l-1)}{m-j} &= 2^{n-2},
\end{align}
where $M=\{m_1,\ldots,m_k\}$.

These formulas express the balancedness of $\ket{D}$ and its $Z$-transforms as precise combinatorial constraints on the admissible weights $m_j$, allowing for a systematic enumeration of all symmetric hypergraph codes with distance $2$.
Armed with these results, the authors performed an exhaustive computer search up to $30$ qubits finding many codes satisfying the binomial conditions, hence achieve distance $2$, with the smallest genuine hypergraph code occurring on $8$ qubits.
However, no codes of distance $3$ or higher were found, and in fact additional arguments show that in the special case of symmetric graph codes (where all hyperedges are of size two), it is impossible to achieve distance greater than $2$.

An interesting scenario arises when one assumes that certain physical qubits are protected from errors~\cite{dong2008quantum}, for example, when some qubits are encoded in more stable degrees of freedom than others, or part of the state is distributed without traversing a noisy channel.
In this case one can construct codes~\cite{Analysis_of_qua_Wagner_2018} by starting directly from a hypergraph $D$ defined on the protected vertices, chosen such that the state $\ket{D}$ is balanced (but not necessarily symmetric).
The unprotected vertices are then added to $D$, and for each edge $e$ of $D$ and each new vertex $i$ in the noisy set $A$, an edge $e \cup \{i\}$ is added to the enlarged hypergraph $H$, while $e$ itself is removed.
The resulting state $\ket{H}$ then includes both the protected and unprotected qubits.
The role of $D$ here is analogous to the $X$-difference hypergraph used in the symmetric case.
The actual code space is then defined in a similar way from the hypergraph basis generated by $\ket{H}$.
While this method can in general generate a large code space, it is not clear whether hypergraph states offer an advange with respect to simple graph states in this regard.

A different approach was considered in~\cite{Quantum_Error_C_Balaku_2017}, which is a more direct generalization of the method in~\cite{schlingemann2001quantum} for graph states.
In this case the code is defined by an isometry
\begin{equation}
v: \Hil \to \mathcal{K}, \qquad \dim \mathcal{K} > \dim \mathcal{H},
\end{equation}
while the noisy channel is defined as in equation~\eqref{eq:noise_chan}.
The code $v$ corrects $\mathcal{E}$ if there exists a recovery operator $\mathcal{R}$ such that
\begin{equation}
    \mathcal{R} \big( \mathcal{E}(v\rho v^\dagger) \big) = \rho, \qquad \forall \rho \in \Hil.
\end{equation}

This condition is translated into the hypergraph setting by constructing the encoding isometry from a given $k$-uniform or general hypergraph $H=(V,E)$.
In $V$ we define two kinds of vertices called input and output vertices whose sets are denoted respectively by $X \subseteq V$ and $Y \subseteq V$.
Consider also a finite abelian group $G$ and call $G^K$ for any $K \subseteq V$ the set of possible assignments $G^K = \{g_K = (g_u)_{u \in K} : g_u \in G\}$.
The encoding isometries $v_H: L^2(G^X) \to L^2(G^Y)$ are then defined from the sets $X,Y$ and constructed from $k$-uniform bicharacter functions of the group $G$, extending the bicharacter method introduced in~\cite{schlingemann2001quantum} for graph states.

Using this method, the authors derive a condition for the error detection based on a homomorphism $F$ encoding neighborhood relations in the hypergraph.
Specifially, $F^K_L: G^L \to G^K$ is defined as
\begin{equation}
    F_L^K(g^L) = \left( \sum_{l \in L} g_l f(l,k)\right)_{k \in K}
\end{equation}
where $f(l,k)$ is 1 when $l$ and $k$ are neighbors (i.e. there is an edge $e \in E$ such that $l,k \in e$) and zero otherwise.
With this, they prove that an error configuration appearing in a subset $E \subset Y$ can be detected by the code $v_H$ if and only if the system
\begin{equation}
    F_I^{X\cup E} (g_{X\cup E}) = 0, \qquad I = Y \setminus E,
    \label{eq:qec_hg_condition_if}
\end{equation}
implies
\begin{equation}
    g_X = 0 \quad \text{and} \quad F_X^E (g_E) = 0.
    \label{eq:qec_hg_condition_then}
\end{equation}

Regarding the actual encoding schemes, one can use generalization of the secret sharing scheme with graph states~\cite{markham2008graph}:
\begin{align}
    \ket{0} &\mapsto \ket{0}_D \otimes \ket{H}, \\
    \ket{1} &\mapsto \ket{1}_D \otimes X^{\otimes n}\ket{H},
\end{align}
which encodes one qubit into $n+1$ qubits.
A second possibility is to exploit stabilizers of hypergraph states, defined in \eqref{eq:hg_stab}, and use the encoding given by the stabilizer formalism~\cite{cleve1997efficient}.

Notably, these encoding schemes can be more gate efficent with respect to the ones using graph states.
Indeed, emulating a $k$-hyperedge in a graph state one needs a complete subgraph on $k$ vertices, requiring $k(k-1)/2$ two-qubit $\CZ$ gates.
By contrast, the single $\CCZ{6}$ gate needed for the hypergraph state can be implemented more efficiently.
Specifically for $n \geq 6$, this yields fewer operations than the graph-state equivalent.
For instance, a $6$-qubit hyperedge requires only $12$ two-qubit gates when implemented directly as $\CCZ{6}$, compared with $15$ for the complete graph.

\section{Higher Dimensional Hypergraph States}
\label{sec:qudits}

A further generalization is obtained by moving beyond qubits to higher dimensional systems, or qudits.
Qudit systems are not only of theoretical interest, but also of practical relevance in experimental platforms such as photons with orbital angular momentum, ion traps, and superconducting devices.
In this setting, hypergraph states can be defined for arbitrary local dimension $d$, giving rise to qudit hypergraph states. 

These higher dimensional hypergraph states thus unify and extend several well-known classes: they contain qudit graph states~\cite{hostens2005stabilizer, bahramgiri2006graph, keet2010quantum} as a special subclass hence overlapping with stabilizer states, and in general they provide a compact combinatorial framework for exploring multipartite entanglement in arbitrary dimension.

Finally, it is worth noting that the hypergraph-state formalism has also been generalized to the continuous-variable (CV) setting, extending cluster states to a non-Gaussian regime and enabling the implementation of nonlinear operations~\cite{moore2019quantum}.

\subsection{Qudit Hypergraph States}

In this section we introduce the notion of \emph{qudit hypergraph states}~\cite{Qudit_hypergrap_Steinh_2017,Qudit_hypergrap_Xiong_2018}, which extend the familiar qubit hypergraph states to arbitrary local dimensions $d$.
The construction is built upon the generalized $d$-dimensional Pauli and Clifford groups~\cite{schlingemann2003cluster}.

For a $d$-dimensional quantum system, let $\{\ket{q}\}_{q=0}^{d-1}$ denote the computational basis.
The natural generalization of the Pauli $X$ and $Z$ operators is
\begin{align}
    Z &= \sum_{q=0}^{d-1} \omega^q \ket{q}\bra{q}, \label{eq:genZ}\\
    X &= \sum_{q=0}^{d-1} \ket{q \oplus 1}\bra{q}, \label{eq:genX}
\end{align}
where $\omega = e^{2\pi i/d}$ is the $d$th root of unity and $\oplus$ denotes addition modulo $d$.
These operators are unitary but in general non-Hermitian, instead they satisfy the condition $X^d = Z^d = I$.
They also obey the following commutation relation
\begin{equation}
    X^m Z^n = \omega^{-mn} Z^n X^m,
\end{equation}
and generate the \emph{Pauli group} in dimension $d$.
It can be easily checked that for $d=2$ they reduce to the familiar qubit Pauli matrices.

Similarly to the qubit case, we can also define the generalized \emph{Clifford group}, as the normalizer of the Pauli group in dimension $d$.
In this case, the Clifford group can be generated by the following operators, sometimes called symplectic operators~\cite{schlingemann2003cluster}:
\begin{align}
    S(\xi,0,0) &= \sum_{q=0}^{d-1} \ket{\xi q}\bra{q}, \label{eq:cliff1}\\
    S(1,\xi,0) &= \sum_{q=0}^{d-1} \omega^{\xi q^2 / 2}\ket{q}\bra{q}, \label{eq:cliff2}\\
    S(1,0,\xi) &= \sum_{q=0}^{d-1} \omega^{-\xi q^2 / 2}\ket{p_q}\bra{p_q}, \label{eq:cliff3}
\end{align}
for $\xi \in \mathbb{Z}_d$, where $\ket{p_q}$ denotes the discrete Fourier transform of the basis state $\ket{q}$.
These operators are unitary whenever $\gcd(\xi,d)=1$, and they will be useful later for the classification of qudit hypergraph states under LU equivalence transformations.

Because of the more complex structure of the Pauli group in dimension $d$, instead of simple hypergraphs the states are labeled by combinatorial objects called \emph{multi hypergraphs}.

\begin{definition}[Multihypergraph]
A multihypergraph $H=(V,E)$ consists of a vertex set $V = \{1,2,\ldots,n\}$, and a multiset $E$ of hyperedges $e \subseteq V$.
\end{definition}

\begin{figure}
    \centering
    \includegraphics[width=0.5\columnwidth]{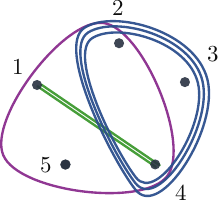}
    \caption{\textbf{Example of a multihypergraph} with $5$ nodes $V=\{1,2,3,4,5\}$ and edges $E = \{\{1,4\}, \{2,3,4\}, \{1,2,4,5\}\}$ with multiplicity $2,3$ and $1$ respectively.
    The corresponding state is $\ket{H} = \CZ_{1,2,4,5} \CZ_{2,3,4}^3 \CZ_{1,4}^2 \ket{+}^{\otimes 5}$.}
    \label{fig:mhg_example}
\end{figure}

Since $E$ is a multiset each hyperedge $e$ may appear with a multiplicity $m_e \in \mathbb{Z}_d$, which encodes the strength of a corresponding multi-qudit interaction (an example is shown in Fig.~\ref{fig:mhg_example}).

The construction of the associated quantum state proceeds by initializing the state in a uniform superposition of all computational basis states, as $\ket{+}^{\otimes n}$, where $\ket{+} = \frac{1}{\sqrt{d}} \sum_{q=0}^{d-1} \ket{q}$.

Then, for each hyperedge $e \in E$ with multiplicity $m_e$, we apply a generalized controlled-phase gate
\begin{equation}
    \CZ_e^{m_e} = \sum_{q_1,\dots,q_{|e|}=0}^{d-1} 
    \omega^{m_e q_1 q_2 \cdots q_{|e|}}
    \ket{q_1,\dots,q_{|e|}}\bra{q_1,\dots,q_{|e|}},
\end{equation}
which imprints phases depending on the joint values of all qudits in the hyperedge, and its multiplicity $m_e$.

\begin{definition}[Multihypergraph states]
Given a multihypergraph $H=(V,E)$, the corresponding \emph{qudit hypergraph state} is
\begin{equation}
    \ket{H} = \prod_{e \in E} \CZ_e^{m_e}\,\ket{+}^{\otimes n}
\end{equation}
\end{definition}

This definition naturally generalizes several well-known classes of states: if all hyperedges have cardinality two, one recovers the family of qudit graph state.
For $d=2$ the construction reduces to ordinary qubit hypergraph states.
Notice that also loops (single-vertex hyperedges) are allowed and correspond to local $Z$ operations, while an empty hyperedge simply contributes an overall phase.

Finally, like in the qubit case (see section~\ref{sec:rew}), the computational basis representation of qudit hypergraph states is also ``equally weighted'', in the sense that each coefficient has the same absolute value~\cite{Qudit_hypergrap_Xiong_2018}.
Specifically they are of the form
\begin{equation}
    \ket{\psi_f} = \frac{1}{d^{n/2}} \sum_{i_1, i_2, \ldots, i_n = 0}^{d-1} 
    \omega^{\, f(i_1,i_2,\ldots,i_N)} \ket{i_1 i_2 \cdots i_n},
    \label{eq:cew}
\end{equation}
where $f: \{0,1,\ldots,d-1\}^n \to \mathbb{Z}_d$ is an integer-valued function defined on the index of the computational basis.
Each basis vector thus appears with equal absolute value $1/\sqrt{d^N}$, but with a phase factor determined by $f$.

\subsection{Stabilizer Formalism for Qudit Hypergraph States}

We have seen in section~\ref{sec:hg_stabilizers} that an elegant feature of qubit hypergraph states is that they can be described as the unique common eigenstate of a commuting family of (generally nonlocal)  operators: the stabilizers.
Qudit hypergraph states admit a similar description, although, just like the generalized Pauli group, the stabilizers in this case are not necessarily Hermitian.
Specifically, let $H=(V,E)$ be a multihypergraph with $n$ nodes and with hyperedge multiplicities $\{m_e\}_{e \in E}$.
For each vertex $i \in V$, we can define the stabilizer operator
\begin{align}
    K_i = X_i \prod_{e \in E: i \in e} \CZ_{e \setminus \{i\}}^{m_e},
    \label{eq:mhg_stabilizer}
\end{align}
where $X_i$ is the generalized $X$ operator acting on qudit $i$, and $\CZ_{e \setminus \{i\}}$ denotes the controlled-phase gate acting on all vertices of the hyperedge $e$ except for $i$.

By construction, the qudit hypergraph state $\ket{H}$ is the unique simultaneous $+1$ eigenstate of all the operators $\{K_i\}_{i \in V}$:
\begin{align}
    K_i \ket{H} = \ket{H}, \qquad \forall i \in V.
\end{align}
Similarly to the qubit case, all the $K_i$ operators commute, therefor the set $\{K_i\}$ generates an Abelian group, called the stabilizer group of $\ket{H}$.
This group has order $d^N$, and, just like in the qubit case, its common eigenbasis is called a \emph{hypergraph basis} whose element can be written as
\begin{equation}
    \ket{H_{\vec m}} = \prod_{i=1}^n Z_i^{-m_i} \ket{H}
\end{equation}
for any choice of the multiplicities $\vec m = (m_1, \ldots, m_n) \in \mathbb{Z}_d^n$.

\subsection{Multipartite Entanglement of Qudit Hypergraph States}

The structure of multipartite entanglement for qudit hypergraph states was studied already in~\cite{Qudit_hypergrap_Steinh_2017} and~\cite{Qudit_hypergrap_Xiong_2018}, and more recently in~\cite{Multipartite_en_Malpet_2022}.
Similarly to the qubit case considered in section~\ref{sec:entanglement}, we are interested in both the characterization of multipartite entanglement and criteria for the identification of LU and SLOCC equivalence classes.

In particular, in~\cite{Qudit_hypergrap_Steinh_2017} the authors studied the latter problem, combining different analytical and numerical techniques.
For any two states $\ket{\psi}$ and $\ket{\phi}$ with subsystems $S = \{1,\ldots,n\}$, a first basic criterion for SLOCC inequivalence comes from the notion of the reduced rank, i.e. the rank of the reduced state $\rho_i = \Tr_{S \setminus i}$.
If the reduced ranks are different, the two states are not equivalent under SLOCC (and therefore neither under LU).

A more refined technique is based on the Minimally Entangled Bases (MEBs)~\cite{lamata2006inductive}.
Consider a state $\ket{\psi} = \sum_{\vec k} \alpha_{\vec k} \ket{k_1,\ldots,k_n}$, and a given subsystem partition under $1|2\ldots n$ that we want to analyze.
The idea is that one can define the matrix
\begin{equation}
    C_{1|2,\ldots,n} = \sum_{\vec k} \alpha_{\vec k} \ket{k_1}\bra{k_2,\ldots,k_n}.
\end{equation}
Now the Singular Value Decomposition (SVD) of $C_{1|2\ldots n}$ will naturally define a subspace of $2,\ldots,n$.
We call a MEB, the basis for this subspace that has the maximum number of product states.

Consider now two states having the same reduced ranks.
One can prove that if they have MEBs with a different number of product vectors then they are inequivalent under SLOCC.
If instead their MEBs are complete product bases then the two are SLOCC equivalent.

\subsubsection{Elementary hypergraph states}

Consider the family of $n$-elementary hypergraph states $\ket{G_n^m}$, generated by a single hyperedge $e$ of cardinality $n$ and arbitrary multiplicity $m$.
In this case the problem simplifies considerably. 
Indeed for an $n$-elementary hypergraph state with hyperedge multiplicity $m$, the rank of any reduced state is given by
\begin{align}
    \mathrm{rank}(\rho) = \frac{d}{\gcd(d,m)}.
\end{align}
Because SLOCC operations cannot change ranks of reduced states, this formula immediately supplies a family of invariants that can distinguish inequivalent entanglement classes.

Another important result is that two $n$-elementary hypergraph states with multiplicities $m$ and $m'$ belong to the same SLOCC class if and only if
\begin{align}
    \gcd(d,m) = \gcd(d,m').
\end{align}
Notably, this gcd criterion implies that prime dimensions collapse all multiplicities into a single SLOCC class, whereas composite dimensions can have multiple inequivalent families.

\subsubsection{Classification of qutrits and ququarts}
Using the MEB technique together with semi-definite programming (SDP) numerical optimizations, in~\cite{Qudit_hypergrap_Steinh_2017} the authors managed to classify the entanglement families for three qudits with $d=3$ and $d=4$.

Dimension $d=3$ is a case of prime dimension, therefore all genuinely entangled elementary hypergraph states are SLOCC equivalent.
In this case the LU classification reveals two inequivalent families.
One is the graph state family, represented by the three-qutrit GHZ state and its LU equivalents, the other is the family of 3-elementary hypergraph states, characterized by a single 3-hyperedge.
While the two families are not LU-equivalent, one can find explicit invertible local maps connecting the two, proving that there is a single SLOCC class.

The case of three ququarts already exhibits a much richer structure because composite dimension allows multiple gcd values.
In total, Steinhoff \emph{et al.}~\cite{Qudit_hypergrap_Steinh_2017} were able to identify six LU classes, and five distinct SLOCC families.

\subsection{Bounds on Geometric Entanglement in Qudit Hypergraph States}

Recently it has been shown~\cite{Multipartite_en_Malpet_2022} that entanglement in qudit hypergraph states can be quantified and bounded directly from their structural features (dimension, multiplicity, and edge size).
In particular these results provide bounds and analytical values for the multipartite version of geometric entanglement introduced in section~\ref{sec:geom_ent}.

A first set of results concerns elementary hypergraph states.
As before, this class of hypergraphs allows for significant simplifications and their entanglement content can be fully characterized analytically.

Let $d = \prod_{i\in J} p_i^{l_i}$ be the prime factorization of the local dimension and consider an elementary state $\ket{G^{m}_n}$ on $n$ qudits and multiplicity $m$, such that
\begin{equation}
    \gcd(m,d) = \prod_{i\in J} p_i^{k_i}.
\end{equation}

Then, the geometric multipartite entanglement, as defined in equation~\eqref{eq:geom_ent} is given by

\begin{widetext}
\begin{equation}
    E_M(\ket{G^{m}_n}) = 1 - \prod_{i\in J} \left[
        1 - (1-\delta_{k_i,l_i})
        \left(1-\frac{1}{p_i}\right)^{n-1}
        \sum_{j=0}^{l_i-k_i-1} 
        p_i^{-j} \binom{n+j-2}{j}
    \right].
    \label{eq:geom_hgele}
\end{equation}
\end{widetext}

This explicit expression allows for several insights.
First, two elementary states with multiplicities $m$ and $m'$ such that $\gcd(m,d) = \gcd(m',d)$, display exactly the same entanglement.
This is expected since we have seen that they are SLOCC equivalent in this case~\cite{Qudit_hypergrap_Steinh_2017}.

Moreover, in prime dimensions, the dependence on multiplicity disappears altogether, and the expression~\eqref{eq:geom_hgele} reduces to the particularly simple form
\begin{equation}
  E_M(\ket{G^{m}_n}) = \frac{(d-1)^{n-1}}{d^{n-1}}.
\end{equation}

In composite dimensions, the situation is more intricate: entanglement decreases monotonically with the number of qudits, and the specific value depends on the arithmetic relation between multiplicity and $d$.
In particular, the minimally entangled states are obtained for multiplicity $d/\mathrm{lpf}(d)$, where $\mathrm{lpf}(d)$ is the least prime factor of $d$, while the maximally entangled ones correspond to multiplicities coprime with $d$.
This yields the bounds
\begin{equation}
  E_M\left(\ket{G^{\,d/\mathrm{lpf}(d)}_n}\right) 
      \leq E_M(\ket{G^{m}_n}) 
      \leq E_M(\ket{G^{1}_n}) .
\end{equation}
since $1$ is coprime with any number.

Consider now more general hypergraph states $\ket{H_{n, k_{\max}}}$, where $k_{\max}$ is the largest hyperedge cardinality.
Starting from such a state $\ket{H_{n,k_{\max}}}$, choosing an arbitrary bipartition, and using only local operations (with respect to that bipartition) one can always reduce it to a state of the form
\begin{equation}
  \ket{G^\mu_\kappa}\otimes \ket{q_1}\otimes \cdots \otimes \ket{q_{n-\kappa}},
  \label{eq:eleform}
\end{equation}
with $\ket{G^\mu_\kappa}$ an elementary hypergraph state of $2 \leq \kappa \leq k_{\max}$ qudits and some multiplicity $\mu$.
Since entanglement cannot increase under local transformations, the entanglement of the original state must be at least as large as that of the least entangled elementary state.

This allows us to lower bound its multipartite entanglement as
\begin{equation}
    E_M(\ket{H_{n,k_{\max}}}) \geq E_M\left(\ket{G^{d/\mathrm{lpf}(d)}_{k_{\max}}}\right),
\end{equation}
since we know from~\eqref{eq:geom_hgele} that the elementary states with multiplicity $d/\mathrm{lpf}(d)$, are the minimally entangled ones.

Thus, any connected hypergraph state is guaranteed to possess at least the multipartite entanglement of an elementary state with the same largest edge size and multiplicity $d/\mathrm{lpf}(d)$. For prime dimensions, the expression simplifies dramatically to
\begin{equation}
  E_M(\ket{H_{n,k_{\max}}}) 
    \geq \frac{(d-1)^{k_{\max}-1}}{d^{k_{\max}-1}}.
\end{equation}

Tighter bounds can be obtained, using the same technique, for certain families of hypergraph, whose particular structure constraints their elementary graph form~\eqref{eq:eleform}.

\subsection{Hypergraph States in Continuous Variables}

Quantum systems can be broadly categorized into two paradigms: discrete-variable (DV) and continuous variable (CV) systems~\cite{lloyd1999quantum, weedbrook2012gaussian}. 
While DV systems encode information in finite dimensional Hilbert spaces spanned by discrete computational basis states, CV systems utilize infinite dimensional Hilbert spaces. 

The prototypical example is a bosonic mode, such as an electromagnetic field mode in quantum optics~\cite{mandel1996optical,leonhardt1997measuring}.
The system is characterized by a pair of canonically conjugate observables, typically denoted as the position-like $\hat q$ and momentum-like $\hat p$, called phase space quadratures, which can be expressed in terms of the creation and annihilation operators $a_i^\dagger,a_i$ for each mode $i$ of the field as
\begin{align}
    \hat q_i = &\frac{(a_i^\dagger+a_i)}{2}\\
    \hat p_i = i&\frac{(a_i^\dagger-a_i)}{2}\,,
\end{align}
and obey the canonical commutation relation $[\hat q_i, \hat p_j] = \frac{i}{2} \delta_{ij}$.

The computational basis states are typically chosen as the eigenstates of these quadrature operators, $\ket{q}_{q_i}$ and $\ket{p}_{p_i}$.
Because they exist in a continuous spectrum, these eigenstates are unphysical on their own, but they form a complete basis for any state $\ket{\psi}$.

A very useful alternative representation for CV states is the Wigner function $W(q_i,p_j)$, a real function on phase space coordinates~\cite{leonhardt1997measuring}.
For a single mode described by the phase space $(q,p)$ the Wigner function associated to a given hermitian operator $A$, is defined as
\begin{equation}
    W_A(q,p) = \frac{1}{2\pi}\int_{-\infty}^\infty dw \Braket{q-\frac{w}{2} | A | q+\frac{w}{2}} e^{ipw}\, .
\end{equation}
For well defined state $\rho$ the Wigner function is normalized $\int dq dp W\rho(q,p) = 1$, but can take both positive and negative values, i.e. it is a quasi-probability distribution.
An important property of the Wigner function is that we can write the Hilbert-Schmidt product of Hermitian operators $A,B$ directly in terms of $W(q,p)$
\begin{equation}
    \Tr(AB) = W_A(p,q) = \int_{-\infty}^\infty dq dp W_A(p,q)W_B(p,q)\, .
\end{equation}
and in general describe all the dynamics of CV quantum systems in terms of Wigner functions.

\paragraph*{Phase Space Representation and Gaussian States}

A crucial distinction in CV quantum information is between \textit{Gaussian} and \textit{non-Gaussian} states and operations~\cite{eisert2002conditions,walschaers2021non}.
Gaussian States have Wigner functions that are entirely positive and take the form of multidimensional Gaussian distributions~\cite{hudson1974wigner}. 
Coherent states, squeezed vacuum states, and thermal states are all Gaussian. 
The continuous-variable cluster states typically used in MBQC are highly entangled multimode Gaussian states, such as cluster states~\cite{menicucci2006universal,gu2009quantum}.
In the same spirit Gaussian operations are transformations that map Gaussian states to Gaussian states. 
They correspond to Hamiltonians that are at most quadratic in $\hat{q}_i$ and $\hat{p}_i$.
A simple example is the displacement operator defined by $D_{q_i}(s) = e^{is\hat p_i}$ or $D_{p_i}(s) = e^{is\hat q_i}$ which represents a translation in $q_i$ and $p_i$ respectively.
In general, similarly to the qudit case, we can decompose any Gaussian transformation using the following three basic operations:
\begin{align}
    \label{eq:CV_rot}
    &\text{Rotations:} &&R_i(\theta) = e^{i\frac{\theta}{2} ({\hat q_i}^2+{\hat p_i}^2)}\,,\\
    \label{eq:CV_squeeze}
    &\text{Squeezing:} &&S_i(\xi) = e^{-i\frac{\xi}{2}(\hat q_i\hat p_i + \hat p_i\hat q_i)}\,,\\
    \label{eq:CV_CZ}
    &\text{Control-Phase:} &&\CZ_{ij}(w) = e^{iw\,\hat q_i \hat q_j}\,.
\end{align}
The first simply represents a rotation in phase space, while the second act as a ``squeezing'' of the corresponding Wigner function in a specific direction and models common nonlinear optical effects like parametric down-conversion.
The Control-Z operation instead models interactions between different nodes (e.g. an optical beamsplitter).

Gaussian measurements are then defined as those measurements represented by a POVM proportional to a Gaussian state.
For instance, a prototypical example is the homodyne detection, which measures the projection in a specific direction in the phase space.

While Gaussian states and operations are relatively straightforward to generate and manipulate experimentally, they are subject to a CV analogue of the Gottesman-Knill theorem~\cite{mari2012positive}. 
Standard Gaussian operations alone can be efficiently simulated classically and are therefore insufficient for universal quantum computation.
To achieve a universal gate set one must introduce at least one non-Gaussian element.
This can be, for instance, a cubic phase gate (generated by a Hamiltonian proportional to $\hat{q}^3$), or non-Gaussian measurements like photon counting~\cite{walschaers2018tailoring}. 

\subsubsection{CV Hypergraph states}
\label{sec:cv}

Graph states can be defined for CV system by using the $\CZ_{ij}(w)$ operation, defined in~\eqref{eq:CV_CZ}.
\begin{definition}[CV graph state]
    Given a graph $G=(V,E)$, and a set of edge-weights $W = \{w_{ij}\}_{(i,j)\in E}$ the corresponding state is defined as:
\begin{equation}
    \ket{G, W} = \prod_{ij \in E} e^{i w_{ij} \hat q_i \hat q_j} \ket{0}_p\,.
\end{equation}
\end{definition}
States of this form are by definition Gaussian states, since $\CZ_{ij}(w)$ is a Gaussian operation.
Often the state is defined only on the graph $\ket{G}$, without weights, in that case we just impose $w_{ij} = 1$ for all edges.
These states can be uniquely described by their stabilizers operators:
\begin{equation}
    K_i(s) = D_{q_i}(s) \sum_{j \in N(i)} D_{p_j}(s)
\end{equation}
with the usual property that $K_i(s)\ket{G} = \ket{G}$ for all $x \in \Real$.
Alternatively one can also define the nullifier operators
\begin{equation}
    H_i = \hat p_i - \sum_{j \in N(i)} q_j
\end{equation}
for which $H_i\ket{G} = 0$, and represent the Lie algebra corresponding to the stabilizer group.

Hypergraph states instead, similarly to their discrete counterpart, embed non-Gaussianity (like $\hat q_i \hat q_j \hat q_k$ nonlinearities) directly into the state, which in principle, provides a resource capable of universal quantum computation even when the processing is implemented exclusively via Gaussian operations and measurement~\cite{moore2019quantum}.

In the case of hypergraphs, vertices are still represented by zero-momentum eigenstates, $\ket{0}_p$, while edges are now generated by a generalized controlled-Z operator:
\begin{equation}
    \CZ_{i_1, \dots, i_k}(w) = e^{i w\,\hat q_{i_1} \dots \hat q_{i_k}}\,.
\end{equation}

\begin{definition}[CV hypergraph state]
The state corresponding to the hypergraph $H=(V,E)$, together with a set of edge-weights $W=\{w_e\}_{e\in E}$ is defined as:
\begin{equation}
    \ket{H, W} = \prod_{e \in E} \CZ_{e}(w_e) \ket{0}_p^{\otimes n}\,.
\end{equation}
\end{definition}

These states are stabilized by specific operators of the form
\begin{equation}
    K_i(s) = D_{q_i}(s)\prod_{e \in E(i)} \CZ_{e \setminus i}(s)\,,
\end{equation}
where we are using the same notation as in~\eqref{eq:hg_stab}.
The corresponding generalized nullifiers are
\begin{equation}
    H_i = \hat p_i - \sum_{e \in E(i)} \prod_{j \in e\setminus i}\hat q_j\,.
\end{equation}
It can be easily checked that all the $H_i$ commute and any other nullifier $N$ such that  $N \ket{H} = 0$ can be expressed a linear combination of them $N = \sum_i \alpha_iH_i$.

\subsubsection{Gaussian Operations}

The framework of CV hypergraph states provides a simple way to track the evolution of a quantum state $\ket{H}$ under Gaussian operations, directly through modifications to this underlying hypergraph geometry and the weights $w_e$.
The operations that allow this description are summarized in table~\ref{tab:cv_ops}.
Following~\cite{vandre2025graphical}, to simplify the notation, we will denote as $(E;W) = \{(e,w_e): e \in E\}$ the set of the tuples of edges with their respective weight, and when doing the union of two sets $(E;W) \cup (E';W')$, we will add the weights whenever an edge belongs to both set.

\paragraph*{Displacements:} 
Consider a simple displacement operation. 
When a momentum displacement $D_{p_i}(s) = e^{is\hat{q}_{i}}$, is applied to a specific mode $i$, the operation either introduces a new single-vertex hyperedge $\{i\}$ with a weight of $s$, or simply increases the weight of that hyperedge by $s$ if it is already present in the graph.
\begin{equation}
    (E'; W') = (E; W) \cup (\{i\}; s)\,.
\end{equation}

In contrast a position displacement $D_{q_i}(s) = e^{-is\hat{p}_{i}}$ generates or modifies hyperedges among the neighboring modes:
\begin{equation}
    (E'; W') = (E; W) \cup (A(i); \{\bar w_e\}), 
\end{equation}
where $\bar{w}_e = -s w_{e\cup \{i\}}$ for each $e \in A(i)$.

\paragraph*{Squeezing:}
A squeezing operation $S_i(\xi)$ applied to mode $i$, will scale the weights of all hyperedges connected to that mode by a factor of $e^{-\xi}$.
The resulting graph takes the form:
\begin{equation}
    (E'; W') = (E; W) \cup (E(i); \{\bar w_e\})\,,
\end{equation}
where $\bar{w}_e = (e^{-\xi} - 1) w_{e}$ for each $e \in E(i)$.

\paragraph*{Rotation:}
A rotation operation $R_i(\theta)$ preserves the hypergraph state strictly when the rotation angle satisfies $\theta = n\pi$. 
Under this condition, the graph remains entirely unchanged if $n$ is even. 
If $n$ is odd, the structure persists, but the weights of all hyperedges in $E(i)$ are inverted:
\begin{equation}
    (E'; W') = (E; W) \cup (E(i); \{\bar w_e\})\,,
\end{equation}
where $\bar{w}_e = ((-1)^{n} - 1) w_{e}$ for each $e \in E(i)$.
The result for a general angle $\theta$ instead does not have a simple graphical interpretation, and produces a state that is not necessarily a hypergraph state (see~\cite{vandre2025graphical}).

\paragraph*{Control-Z}
A control-Z gate $\CZ_{i,j}(w)$, from a graphical standpoint, simply adds an edge of coupling strength $w$ between modes $i$ and $j$, or shifts its weight if the modes are already linked:
\begin{equation}
    (E'; W') = (E; W) \cup (\{i, j\}; w)
\end{equation}

\paragraph*{Measurements:}
Finally, Gaussian measurements alter the topology of the state by severing connections. 
Projecting a mode onto a position quadrature eigenstate, $\ket{q}_{q_i}\bra{q}_{q_i}$, collapses the state such that mode $i$ is disconnected from the hypergraph. 
The effect of this measurement also affects its former neighbors, scaling the weights of the adjacent hyperedges by a factor of $q$:
\begin{equation}
    (E'; W') = (E \setminus E(i); W) \cup (A(i); \{\bar{w}_e\})\,,
    \label{eq:position_measure_CV}
\end{equation}
where $\bar{w}_e = q w_{e \cup a}$
Performing a momentum measurement, $\ket{p}_i\bra{p}_i$, similarly removes mode $i$ from the hypergraph. 
However, the subsequent update to the remaining graph is more mathematically involved, effectively requiring an integration over all possible edge weights to determine the new correlations:
\begin{multline}
    \ket{p}_{p_i}\bra{p}_{p_i}\ket{H} = \ket{p}_{i} \int_{-\infty}^{\infty} dx \, e^{-ipx} \prod_{e \in A(i)} C_{e}(x w_{e \cup i}) \\
    \prod_{e' \in E \setminus E(i)} C_{e'}(w_{e'}) \ket{0}_{p}^{\otimes |V|-1}\,.
    \label{eq:momentum_measure_CV}
\end{multline}

\begin{table}[htp]
\centering
\begin{tabular}{llp{5cm}}
\textbf{Operation} & \textbf{Symbol} & \textbf{Hypergraph description} \\ 
\midrule
Control-Z & $CZ_{e}(w)$ & Adds a hyperedge $e$ with weight $w$, or increases its weight by $w$.\\
$q$-Displacement & $D_q(s)$ & Adds an edge $\{i\}$ with a weight of $s$, or increases its weight by $s$. \\
$p$-Displacement & $D_q(s)$ & Generates or modifies the weight of edges in adjacent edges $e \in A(i)$ by $-s w_{e\cup \{i\}}$. \\
Squeezing & $S_i(\xi)$ & Multiplies by $e^{-\xi}$ the weights of all edges connected to $i$.\\
Rotation & $R_i(n\pi)$ & For odd $n$ invert the weights of all edges in $E(i)$, for even $n$ does nothing. \\
$q$-Measurement & $\ket{q}_{q_i}\bra{q}_{q_i}$ & Disconnects $i$ and scales the adjacent vertices by $q$ \\ 
$p$-Measurement & $\ket{p}_{p_i}\bra{p}_{p_i}$ & Disconnects $i$ and updates the state as in~\eqref{eq:momentum_measure_CV}.  \\ 
\bottomrule
\end{tabular}
\caption{Gaussian operations in CV hypergraph states and their description in terms of the hypergraph representation.}
\label{tab:cv_ops}
\end{table}

\subsubsection{Shaping Hypergraphs with Gaussian Measurements}
\begin{figure}
    \centering
    \includegraphics[width=0.9\columnwidth]{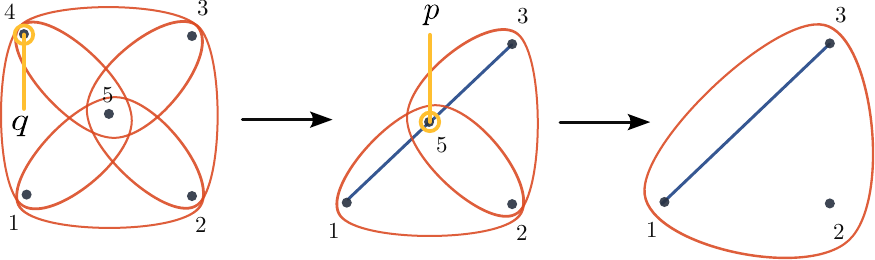}
    \caption{Representation of the protocol for the application of the cubic phase gate protocol using a cell of a $3$-cluster state.
    By performing first a position $q$ measurement and then a momentum $p$ measurement the $3$-edge is moved on the vertices $1,2,3$, effectively applying the cubic gate on a new set of vertices.}
    \label{fig:3phase_gate}
\end{figure}
Of particular interest are the $k$-uniform hypergraph states $\ket{H^k_n}$.
An important class, relevant for MBQC, is represented by a 3-uniform hypergraph, also called ``3-cluster state'', which introduces a nonlocal cubic nonlinearity proportional to $\hat q_i \hat q_j \hat q_k$~\cite{moore2019quantum}.

In~\cite{moore2019quantum}, the author proves that such a 3-uniform hypergraph, utilizing only a Gaussian measurement scheme, is sufficient to generate a 3-edge between three vertices that did not previously share one (effectively ``teleporting'' the nonlinearity). 

This protocol relies heavily on how Gaussian measurements reduce the complexity of hyperedges, specifically how position $q$ and momentum $p$ measurements convert $k$-hyperedges to $(k-1)$-hyperedges.
As an example consider a cell with a configuration like the first step in Fig.~\ref{fig:3phase_gate}.
The protocol consists in performing a position measurement on the upper right vertex, which, according to~\eqref{eq:position_measure_CV} gets disconnected from the graph, and creates the two $2$-edges as shown in the second step in Fig.~\eqref{fig:3phase_gate}.
Then a momentum measurement is performed on the central vertex, which detaches it and produces a new $3$-edge between the remaining vertices, as can be checked by using~\eqref{eq:momentum_measure_CV}.

It can be shown~\cite{moore2019quantum} that as a consequence of this result the cubic phase gate, a strictly non-Gaussian operation, can be implemented on any arbitrary input state connected to a $3$-cluster using exclusively this kind of Gaussian measurements with the help of Gaussian ancillary modes.

Moreover, by systematically applying $\hat q$ measurements to specific central vertices in the alternating 3-cluster geometry, the 3-edges can be selectively reduced to standard 2-edges cluster state, which are known to be universal for Gaussian operations. 
Hence these operations, satisfy both the non-Gaussian (cubic phase) and Gaussian operational requirements purely through homodyne detection.

\section{Conclusions}

Quantum hypergraph states constitute a natural generalization of graph states, obtained by extending controlled-phase interactions to multi-qubit hyperedges.
This seemingly simple modification yields a class of states with substantially richer entanglement properties, and a stabilizer structure that goes beyond the Pauli group.
As such, hypergraph states provide both a conceptual framework for analyzing genuinely multipartite entanglement and an interesting resource for quantum computation and communication.

This review has focused in particular in the complexity of their entanglement structure.
Even for relatively small systems, hypergraph states display LU- and SLOCC-inequivalent classes that cannot be reduced to graph states, thus expanding the landscape of multipartite entanglement families.
Tools such as geometric entanglement, multipartite negativity, and stabilizer-based witnesses have proven effective in characterizing these states, while purification protocols demonstrate that their nontrivial entanglement can be distilled in noisy settings.
Moreover, randomized hypergraph models provide insights into their robustness under decoherence and stochastic gate application, with nontrivial phenomena such as entanglement sudden death and rebirth.

From a computational standpoint, hypergraph states offer distinct advantages.
In measurement-based quantum computation, families of 3-uniform states act as universal resources while requiring only Pauli measurements, thereby reducing the need for adaptive measurement bases.
Their intrinsic nonlocal non-stabilizer nature makes them a possible resource for quantum computing.
Recent work quantifying this resource through stabilizer entropies, robustness measures, and entropic monotones has begun to clarify their role within the broader resource-theoretic approach to quantum computation.

Hypergraph states also serve as a fertile ground for foundational investigations. Their generalized stabilizer structure enables the derivation of Bell-type and Hardy-type inequalities, which reveal contextuality and genuine multipartite nonlocality.
Notably, these inequalities exhibit exponential violations with system size and remain robust to particle loss.

Finally, the framework extends naturally to higher-dimensional systems, through hypergraphs with edge multiplicity or weights, and generalize $\CZ$ or C-phase gates.
Qudit and CV hypergraph states inherit many of the entanglement and nonclassical features of their qubit counterparts, while presenting a richer entanglement structure and promising applications in error correction and coding.
These results indicate that the hypergraph construction is not tied to qubit systems, but provides a general formalism adaptable to diverse quantum architectures.

We hope that future work in this area will continue to further elucidate their structure and classification, their relevance for quantum computation, and investigate their experimental realization in scalable quantum architectures, consolidating their role in both the foundational and applied aspects of quantum information science.

\section*{Acknowledgments}
CM acknowledges funding from MUR PRIN (Project 2022SW3RPY).
DB acknowledges support by Deutsche Forschungsgemeinschaft
(DFG, German Research Foundation) under Germany’s
Excellence Strategy – Cluster of Excellence Matter and
Light for Quantum Computing (ML4Q) EXC 2004/1 –
390534769.
DP acknowledges funding from MUR PRIN (Project 2022LCEA9Y).

\section*{List of Abbreviations and Symbols}

\begin{itemize}[align=left,leftmargin=2.0cm,labelwidth=!,nosep,itemsep=0.1cm]

\item[] \textbf{Abbreviations}

\item[CV] Continuous Variables.
\item[EPC] Edge-Pair Complementation.
\item[GME] Genuinely Multipartite Entangled.
\item[GMN] Genuine Multipartite Negativity.
\item[GMNL] Genuine Multipartite Nonlocality.
\item[LC] Local Clifford.
\item[LME] Locally Maximally Entanglable states.
\item[LOCC] Local Operations with Classical Communication.
\item[LU] Local unitary.
\item[MBQC] Measurement-Based Quantum Computation.
\item[MEB] Minimally Entangled Basis.
\item[NPT] Non-positive Partial Transpose.
\item[PPT] Positive Partial Transpose.
\item[QEC] Quantum Error Correction.
\item[REW] Real Equally Weighted states.
\item[RHS] Randomized Hypergraph States.
\item[SLOCC] Stochastic Local Operations with Classical Communication.
\item[SRE] Stabilizer Reny\'i Entropy.
\item[STAB] Stabilizer states set.
\\

\item[] \textbf{Hypergraphs}

\item[$A\sdiff B$] Symmetric difference of sets $A$ and $B$, $A \sdiff B = (A\cup B) \setminus (A \cap B)$.
\item[$H_1 \sdiff H_2$] Symmetric difference of hypergraphs, $V=V_1=V2$ and $E=E_1 \sdiff E_2$.
\item[$N(i)$] Neighbors of a node $i$ in a hypergraph: $N(i)=\{j : (i,j)\in E\}$.
\item[$E(i)$] Adjacent edge set of a node $i$ in a hypergraph: $E(i)=\{e \in E: i \in e\}$.
\item[$A(i)$] Adjacency set of a node $i$ in a hypergraph: $A(i)=\{e\setminus\{i\} : e \in E(i)\}$.
\item[$\Delta(i)$] Degree of a vertex $i$ (cardinality of its neighboring set).
\item[$\bar{\Delta}(H)$] Average vertex degree of the hypergraph $H$.
\item[$F \preceq H$] $F$ is a spanning sub-hypergraph of $H$ (same vertices but only a subset of the edges of $H$).
\item[$H_n^k$] Generic $k$-uniform hypergraph with $n$ nodes.
\item[$S_k$] Star graph with $n$ nodes, where a single central node is connected to all the others.
\item[$\Comp_n^k$] Complete $k$-uniform hypergraph with $n$ nodes ( containing all hyperedges of cardinality $k$).
\item[$\Comp^k$] Complete graph with $n$ nodes ($\Comp_n = \Comp_n^2$). 
\item[$\Comp_n^{\vec k}$] Complete hypergraph with $n$ nodes, (containing all hyperedges of cardinalities $\vec k = (k_1,\ldots,k_n)$).\\

\item[] \textbf{Operations and gates}

\item[$\Pauli_n$] Pauli group on $n$ qubits.
\item[$\Cliff_k$] $k$-th level of the Clifford hierarchy.
\item[$\Cliff$] Clifford group.
\item[$LU_n$] Local unitary group.
\item[$X_i, Y_i, Z_i$] Pauli (or generalized Pauli) operators on node $i$.
\item[$H, S$] Hadamard and Phase gates.
\item[$\CZ$] Control Z gate on two qubits.
\item[$\CNOT$] Control NOT gate on two qubits.
\item[$\CCZ{k}$] Multi-controlled Z and multi-controlled NOT gates on $k$ qubits.
\item[$\CCNOT{k}_{S,t}$] Multi-controlled NOT gates on $k$ qubits, where $S$ is the control and $t$ is the target.
\item[$\CZ_e$] Multi-controlled Z applied to and hyperedge $e$.
\item[$\CZ_e^{m_e}$] Qudits multi-controlled Z applied to and hyperedge $e$ with multiplicity $m_e$.
\item[$\CZ_e(w)$] Continuous variable multi-controlled Z applied to and hyperedge $e$ with weights $w$.
\item[$\mathrm{C}_e(\phi)$] multi-controlled phase gate with phase $\phi$ applied to and hyperedge $e$.
\item[$\sqrt{P}^{\pm}$] Square root of a Pauli operator $P$, with eigenvalues $1$ and $\pm i$
\item[$\sqrt{\CZ_e}^{\pm}$] Square root of a $\CZ_e$ gate, $\sqrt{\CZ_e}^{\pm} = \eye - (1 \mp i) \ket{1\ldots 1}\bra{1\ldots 1}$.
\end{itemize}

\bibliographystyle{unsrtnat}
\bibliography{refs}

\end{document}